\documentclass{ieeeaccess}
\usepackage{stfloats}
\usepackage{cite}
\usepackage{amsmath,amssymb,amsfonts}
\usepackage{graphicx}
\usepackage{textcomp}
\usepackage[utf8]{inputenc}
\usepackage{algorithm}
\usepackage[noend]{algpseudocode}
\algrenewcommand\algorithmicrequire{\textbf{Input:}}
\algrenewcommand\algorithmicensure{\textbf{Output:}}
\usepackage{array}
\usepackage{url}
\usepackage{graphics}
\usepackage{verbatim}
\usepackage{siunitx}
\usepackage{booktabs, makecell}
\ifCLASSOPTIONcompsoc
    \usepackage[caption=false, font=normalsize, labelfont=sf, textfont=sf]{subfig}
\else
    \usepackage[caption=false, font=footnotesize]{subfig}
\fi
\usepackage{nomencl}
\usepackage{etoolbox}
\usepackage{array,multirow}
\usepackage{color, colortbl}
\usepackage{soul}
\usepackage{mathtools}
\usepackage{mdwtab}
\makeatletter
\newcommand*{\rom}[1]{\expandafter\@slowromancap\romannumeral #1@}
\makeatother
\usepackage{float}
\graphicspath{{./Figures/}}
\usepackage{comment}
\usepackage{gensymb} 
\usepackage{bm} 
\usepackage{cleveref} 
\def\BibTeX{{\rm B\kern-.05em{\sc i\kern-.025em b}\kern-.08em
    T\kern-.1667em\lower.7ex\hbox{E}\kern-.125emX}}
   
\begin{document}
\history{Date of publication xxxx 00, 0000, date of current version xxxx 00, 0000.}
\doi{10.1109/ACCESS.2023.DOI}

\title{Adopting Dynamic VAR Compensators to Mitigate PV Impacts on Unbalanced Distribution Systems \vspace{-0.07in}}
\author{\uppercase{Han~Pyo~Lee}\authorrefmark{1}, \IEEEmembership{Student Member, IEEE}, 
\uppercase{Keith~DSouza}\authorrefmark{2}, \IEEEmembership{Member, IEEE},
\uppercase{Ke~Chen}\authorrefmark{3}, \IEEEmembership{Member, IEEE},
\uppercase{Ning~Lu}\authorrefmark{1}, \IEEEmembership{Fellow, IEEE}, and \uppercase{Mesut~Baran}\authorrefmark{1}, \IEEEmembership{Life Fellow, IEEE}}
\address[1]{North Carolina State University, Raleigh, NC 27695 USA (e-mail: hlee39, baran@ncsu.edu)}
\address[2]{ComEd (an Exelon company), Oakbrook Terrace, IL 60181, USA}
\address[3]{Duke Energy, Raleigh, NC 27601, USA}
\tfootnote{This work was supported in part by the U.S. Department of Energy’s Office of Energy Efficiency and Renewable Energy (EERE) under the Solar Energy Technologies Office Grant DE-EE0008770, and in part by the Duke Energy. \vspace{-0.2in}}

\markboth
{H. P. Lee \headeretal: Adopting DVCs to Mitigate PV Impacts on Unbalanced Distribution Systems}
{H. P. Lee \headeretal: Adopting DVCs to Mitigate PV Impacts on Unbalanced Distribution Systems}
\corresp{Corresponding author: Mesut~Baran (e-mail: baran@ncsu.edu).}

\begin{abstract}
The growing integration of distributed energy resources into distribution systems poses challenges for voltage regulation. Dynamic VAR Compensators (DVCs) are a new generation of power electronics-based Volt/VAR compensation devices designed to address voltage issues in distribution systems with a high penetration of renewable generation resources. Currently, the IEEE Std. 1547-based Volt/VAR Curve (VV-C) is widely used as the local control scheme for controlling a DVC. However, the effectiveness of this scheme is not well documented, and there is limited literature on alternative control and placement schemes that can maximize the effective use of a DVC. In this paper, we propose an optimal dispatch and control mechanism to enhance the conventional VV-C based localized DVC control. First, we establish a multi-objective optimization framework to identify the optimal dispatch strategy and suitable placement for the DVC. Next, we introduce two supervisory control strategies to determine the appropriate instances for adjusting the VV-C when the operating condition changes. The outlined scheme comprises two primary stages: time segmentation and VV-C fitting. Within this framework, each time segment aims to produce optimized Q-V trajectories. The proposed method is tested on a modified IEEE 123-bus test system using OpenDSS for a wide range of operating scenarios, including sunny and cloudy days. Simulation results demonstrate that the proposed scheme effectively reduces voltage variations compared to the standard VV-C specified in IEEE Std. 1547.
\vspace{-0.1in}
\end{abstract}


\begin{keywords}
DER impact mitigation, Distribution system, Dynamic VAR Compensator (DVC), High Penetration PV, Smart inverter, Volt/VAR control. 
\vspace{-0.1in}
\end{keywords}

\titlepgskip=-22pt

\maketitle

\section{Introduction} \label{sec:introduction}
\PARstart{T}{he} integration of distributed energy resources (DERs), particularly photovoltaics (PVs), into distribution systems \cite{SEIA} poses challenges for voltage regulation. The high penetration of PVs introduces power fluctuations caused by factors like cloud movements, leading to rapid voltage fluctuations. Conventional voltage control devices, such as Voltage Regulators (VRs), are forced to switch frequently in response to these deviations \cite{seguin2016high}, resulting in a shortened device lifespan and an increased risk of premature failure. To address this emerging challenge, Dynamic VAR Compensators (DVCs) are being evaluated as a solution. In addition to resolving voltage regulation issues, DVCs offer potential benefits such as enhancing power losses, mitigating voltage flicker, and reducing voltage imbalances \cite{padullaparti2016advances,AMSC}.

DVCs are power electronics-based reactive power (Q) compensators. While widely used in transmission voltage regulation, their application in distribution system operation is still in its early stages. DVCs offer fast and continuous control of reactive current, making them a suitable complement to capacitor banks and tap changing regulators. In a study conducted by DSouza et al. \cite{dsouza2022maximizing}, it was demonstrated that DVCs effectively mitigate problems such as excessive tap changes and frequent voltage violations caused by variable PV generation. Additionally, DVCs enable precise and rapid power control on a per-phase basis, ensuring that the voltage across the feeder remains within the limits specified by ANSI standards \cite{ANSI}.

\begin{table*}[t]
	\begin{center}
        \vspace{-0.18in}
		\caption{\textbf{A Review of Existing Methods and Our Contributions.}}
        \vspace{-0.1in}
		\label{tab1}
		\centerline{\includegraphics[width=\linewidth, height=0.37\textheight]{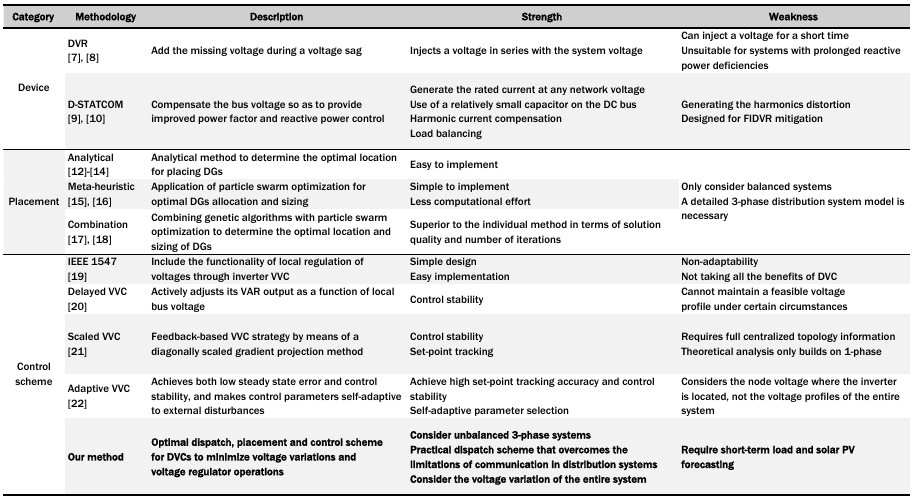}}  
        \vspace{-0.3in}
	\end{center}
\end{table*}

Table \ref{tab1} presents a comprehensive overview of existing methods for addressing the optimal placement and control schemes for DVCs, along with a comparison of their strengths and weaknesses in relation to the proposed approach introduced in this paper.

Power electronics-based voltage regulation devices in distribution systems include Dynamic Voltage Restorer (DVR) \cite{sadigh2012review, monti2020converter}, distribution static synchronous compensator (DSTATCOM) \cite{singh2012design, kumar2014voltage}, and DVC \cite{AMSC}. Among these devices, DVRs are not suitable for systems experiencing prolonged reactive power deficiencies, while DSTATCOM was primarily designed to address Fault-Induced Delayed Voltage Recovery (FIDVR) issues \cite{sun2019review}. In comparison, DVCs are designed to complement existing Volt/VAR Control (VVC) devices by effectively managing feeder voltage within the ANSI-prescribed limits \cite{ANSI}. The primary objective of the DVC is to mitigate voltage violations and fluctuations resulting from intermittent PV outputs, providing necessary voltage boost or reduction.

The challenge in deploying the DVC lies in determining the optimal location(s) for its installation. While various approaches have been proposed for placing devices like distributed generations, most of these approaches rely on analytical methods \cite{wang2004analytical, keane2005optimal, gozel2009analytical}, meta-heuristic techniques \cite{soroudi2012binary, abdi2013application}, or a combination of both \cite{moradi2012combination, tan2013multi}. Although these methods can be adapted for DVC placement, modifications are necessary as they primarily address balanced systems. To address this challenge and facilitate DVC placement, it is crucial to employ a comprehensive 3-phase distribution system model that considers unbalanced system conditions and the operation of per-phase DVCs. Another challenge is to select an appropriate control scheme for maximizing the DVC benefit. Existing approaches rely on the standard Volt/VAR Characteristics (VVAR-C) based local control, as specified in IEEE Std. 1547 \cite{photovoltaics2018ieee}. However, this type of control does not fully exploit the potential advantages offered by DVCs, such as their fast response, usually measured in cycles, and the capability to independently inject corrected reactive power into each phase without dependence on the other phases. Previous literature has presented solutions related to smart inverters including delayed VVC \cite{jahangiri2013distributed}, scaled VVC \cite{zhu2015fast}, and adaptive VVC \cite{singhal2018real}, but these approaches have limitations and drawbacks, as highlighted in Table \ref{tab1}. Moreover, a dedicated tool is needed to optimize the placement and control scheme of DVCs so that distribution planning engineers can plan and deploy these devices more effectively on their systems.

This paper focuses on both the control dispatching and placement problems associated with the adoption of a DVC on a distribution feeder. The paper first proposes a novel DVC dispatching scheme designed to mitigate voltage fluctuations on a feeder with high PV penetration. This scheme adopts a simple dispatch objective, allowing the DVC to react to voltage violations while minimizing excessive voltage regulator operation. The dispatching approach is then integrated into a placement method to identify an optimal location for the DVC, ensuring its effectiveness in voltage regulation. Furthermore, the paper proposes a more practical supervisory control scheme to minimize the frequent dispatches considered in the initial dispatching scheme. This supervisory control scheme periodically adjusts the local VV-C to enable the DVC to adapt to changing operating conditions. This approach addresses the constraints imposed by communication infrastructure limitations, where frequent updates for optimal dispatch (e.g., every 1 minute) are not feasible.

The paper offers three key contributions. Firstly, an optimal DVC dispatching scheme is proposed to minimize voltage variations and reduce the number of VR switching operations. Secondly, a novel method is introduced to identify suitable DVC deployment locations, considering the locational impact on voltage profiles to effectively mitigate voltage variations. Lastly, a supervisory dispatch scheme is proposed to adjust DVC control parameters based on the Q-V trajectory derived from the optimal dispatch. Simulation results demonstrate that the proposed methods surpass standard VVC in reducing voltage variations and regulator operations.

The paper is structured as follows: Section II introduces the optimal dispatch scheme and identifies the suitable location for the DVC. Section III outlines the proposed practical dispatching scheme. Section IV presents simulation results to evaluate the performance of the proposed approach. Finally, Section V concludes the paper.

\vspace{-0.05in}
\section{Optimal DVC Dispatch and Placement}
As discussed in the previous section, the main benefit of utilizing a DVC is the mitigation of voltage variations on a distribution feeder. Voltage variation is directly associated with the degree of voltage fluctuation at each node along the feeder. To maintain voltage variations within the desired limits, typically defined by voltage violation thresholds specified in ANSI standards \cite{ANSI}, utilities employ Line Voltage Regulators (LVRs) and Capacitor Banks (CAPs). The Category I limits, commonly adopted by utilities, range between 0.95 and 1.05 pu. However, with the implementation of Conservation Voltage Reduction (CVR), utilities aim to further reduce voltages on feeders, necessitating tighter control over voltage variations \cite{lee2022iterative}. The DVC proves valuable in achieving this objective by ensuring that voltages remain within a specific target voltage band. This paper considers a voltage band of 0.98 $\sim$ 1.03 pu, as depicted in Fig.~\ref{fig1}.

\Figure[t](topskip=-15pt, botskip=0pt, midskip=0pt)[width=0.48\textwidth, height=0.18\textheight]{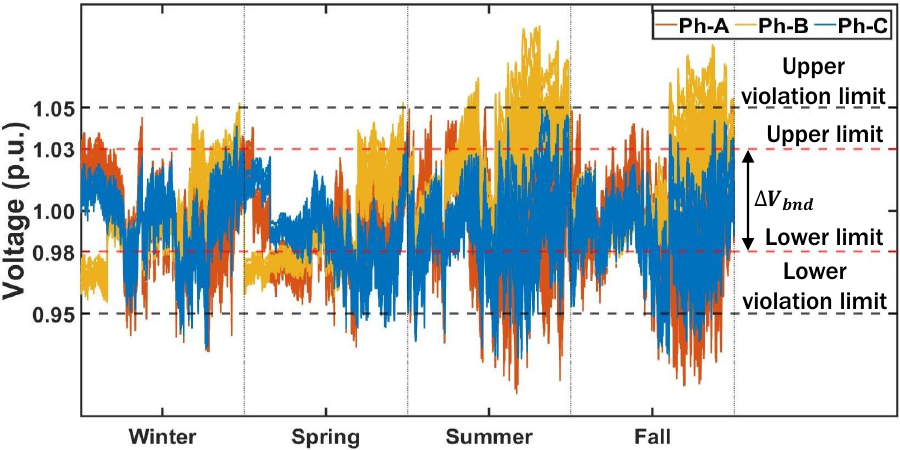} 
{\textbf{Voltage variation limits considered for DVC.} \label{fig1} \vspace{-0.2in}}

\vspace{-0.05in}
\subsection{Dynamic VAR Compensator (DVC)}
The schematic of the novel power electronics-based DVC \cite{AMSC}, which is capable of independently adjusting VAR injection on each phase and exhibits a rapid response time, is depicted in Fig.~\ref{fig2}. These characteristics render the DVC highly efficient in mitigating fast voltage variations and reducing excessive voltage regulator operations resulting from PV systems.

\Figure[hb](topskip=-20pt, botskip=0pt, midskip=0pt)[width=0.48\textwidth, height=0.18\textheight] {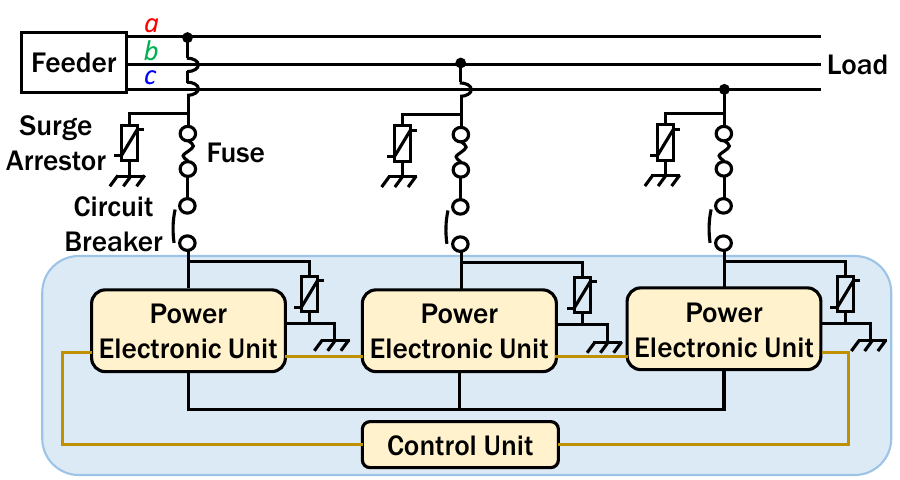}
{\textbf{A Schematic Diagram of the DVC \cite{AMSC}.} \label{fig2} \vspace{-0.2in}}

\subsection{Optimal DVC Dispatch}
The dispatching of a DVC entails determining the desired VAR injection to be provided by the DVC in order to maintain the voltages on the feeder within the specified voltage band, denoted as $\Delta V_{bnd}$. This dispatching problem can be formulated as an optimization problem, where the objective function quantifies the deviation of the node voltages from the $\Delta V_{bnd}$ illustrated in Fig.~\ref{fig1}. Thus, the objective function can be expressed as follows:
\vspace{-0.05in}
\begin{align}
\mathrm{f}^\mathrm{\mu}_\mathrm{j,t} &= \sum_{i\in \mathcal{N},i\notin \mathcal{K}} \biggl( max(V_\mathrm{i,j,t}-V^\mathrm{upper}, 0) \nonumber \\  &\qquad+ max(V^\mathrm{lower}-V_\mathrm{i,j,t}, 0) \biggr), \forall j \in \mathcal{P}, \forall t \in \mathcal{T} \label{eq1} 
\end{align}
where $\mathcal{N}$ represents the set of nodes, $\mathcal{K}$ denotes the set of voltage regulators, $\mathcal{P}$ is the number of phases, $\mathcal{T}$ indicates the scheduling period. $V_\mathrm{i,j,t}$ is the voltage on phase $j$ at node $i$ at time $t$. The lower and upper limits (i.e., $V^{lower}$ and $V^{upper}$) can be set based on voltage variation on the feeder before the DVC is added.  

Due to the potential increase in LVR operations caused by PV intermittency and VAR injection from the DVC, an additional objective function can be introduced to mitigate excessive LVR operation. This objective function is defined as the sum of tap movements of the LVRs, as shown below:
\vspace{-0.05in}
\begin{equation}
\mathrm{f}^\mathrm{\theta}_\mathrm{j,t} = \sum_{k\in \mathcal{K}}|\theta_\mathrm{k,j,t}-\theta_\mathrm{k,j,t-1}|, \forall j \in \mathcal{P}, \forall t \in \mathcal{T} \label{eq2}    
\vspace{-0.05in}
\end{equation}
where $\theta_\mathrm{k,j,t}$ is the tap position of regulator on phase $j$ at node $k$ at time $t$.

By incorporating these objective functions, the problem of optimal dispatch can be formulated as follows:
\vspace{-0.05in}
\begin{IEEEeqnarray}{lll} 
\min_\mathrm{Q_\mathrm{j,t}^\mathrm{inj}} \ &\Bigl( w_\mathrm{\mu}f^\mathrm{\mu}_\mathrm{j,t}+w_\mathrm{\theta}f^\mathrm{\theta}_\mathrm{j,t} \Bigr) \label{eq3}\\
\text{s.t.}  
& 0 \leq |Q_\mathrm{j,t}^\mathrm{inj}| \leq 1, \forall j \in \mathcal{P}, \forall t \in \mathcal{T} \label{eq4}
\vspace{-0.05in}
\end{IEEEeqnarray}
The first objective function, which aims to reduce voltage variation, is assigned higher weights to emphasize its importance. To solve this problem, an iterative search method is used to determine the optimal $Q^\mathrm{inj}$ from the DVC for a given feeder operating condition, considering load and PV levels.

\vspace{-0.05in}
\subsection{DVC Dispatch Performance}
To assess how much the DVC reduced the voltage variations and limited the voltage regulator operations, four performance metrics are used: lower voltage violations ($\mathrm{V}^\mathrm{lower}_\mathrm{out}$), within a target voltage band ($\mathrm{V}_\mathrm{in}$), upper voltage violations ($\mathrm{V}^\mathrm{upper}_\mathrm{out}$), and voltage regulator operations ($\mathrm{Tap}_\mathrm{k}$).
\begin{IEEEeqnarray}{llr}
\mathrm{V}^\mathrm{lower}_\mathrm{out} &= \sum_{\tau\in \mathcal{T_\mathrm{1}}} \tau, \mathcal{T_\mathrm{1}} = \{ t \in \mathcal{T}\ |\ V_\mathrm{t} < V^\mathrm{lower} \} \label{eq5} \\
\mathrm{V}_\mathrm{in} &= \sum_{\tau\in \mathcal{T_\mathrm{2}}} \tau, \mathcal{T_\mathrm{2}} = \{ t \in \mathcal{T}\ |\ V^\mathrm{lower} \leq V_\mathrm{t} \leq V^\mathrm{upper} \} \quad \label{eq6} \\
\mathrm{V}^\mathrm{upper}_\mathrm{out} &= \sum_{\tau\in \mathcal{T_\mathrm{3}}} \tau, \mathcal{T_\mathrm{3}} = \{ t \in \mathcal{T}\ |\ V^\mathrm{upper} < V_\mathrm{t} \} \label{eq7} \\
\mathrm{Tap}_\mathrm{k} &= \sum_{t\in \mathcal{T}}|\theta_\mathrm{k,t}-\theta_\mathrm{k,t-1}|, \forall k \in \mathcal{K} \label{eq8}
\end{IEEEeqnarray}

\subsection{DVC Placement}
\begin{figure}[t]
\centerline{\includegraphics[width=\linewidth, height=0.28\textheight]{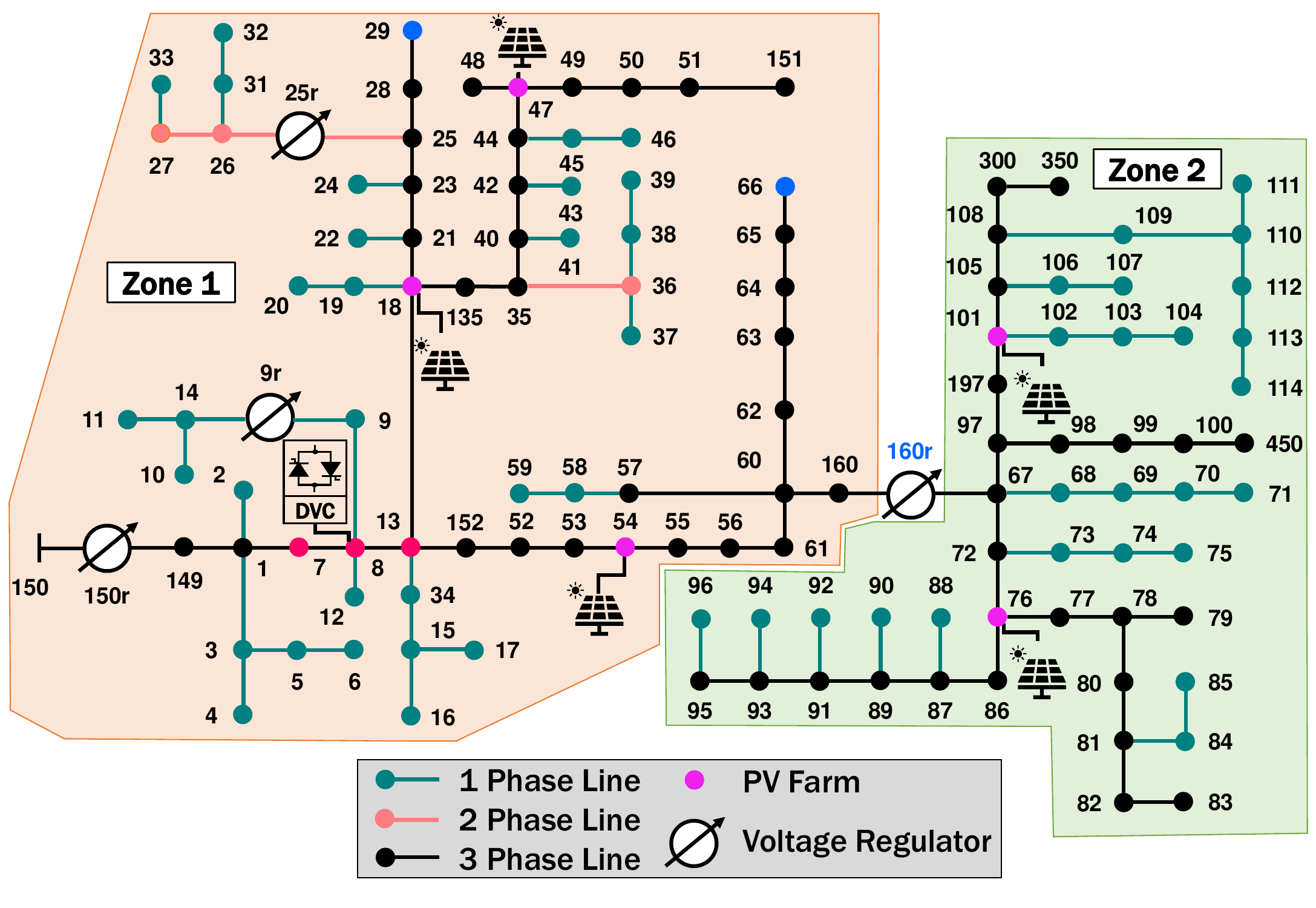}}
\caption{\textbf{IEEE 123 node feeder for test \cite{IEEE123}.}}
\vspace{-0.2in}
\label{fig3}
\end{figure}

Since the DVC injects reactive power, it primarily influences the voltages in the zone in which it is placed. To illustrate this, examine the sample feeder depicted in Fig.~\ref{fig3}. In this system, the LVR (i.e., 160R) on the main feeder divides the feeder into two distinct voltage zones, as indicated in Fig.~\ref{fig3}. Zone 1 represents the first voltage zone (highlighted in orange), while Zone 2 corresponds to the second zone (highlighted in green). 

\begin{figure*}[ht]
\centerline{\includegraphics[width=\textwidth]{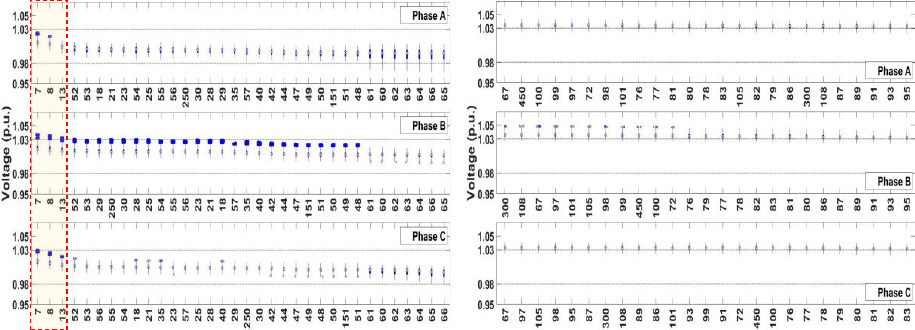}} \vspace{-.2in}
  \subfloat[\label{4a}]{\hspace{.5\textwidth}}
  \subfloat[\label{4b}]{\hspace{.55\textwidth}} 
\vspace{-0.05in}
\caption{\textbf{Node voltage distribution by phase in descending order (a) in Zone 1, (b) in Zone 2.}}
\vspace{-0.1in}
\label{fig4}
\end{figure*}

Time-series power flow simulations are first conducted on the feeder with no DVC, which serves as the base case. Figure~\ref{fig4} shows the phase-wise voltage distribution, sorted in descending order based on average voltage. The figure effectively demonstrates the contrasting voltage variations observed in the two zones. Zone 1 exhibits significantly larger voltage variations compared to Zone 2, mainly due to large PV farms. Furthermore, Zone 1 experiences greater voltage imbalance between phases compared to Zone 2. Consequently, our objective is to examine the effectiveness of the DVC in mitigating voltage variations specifically within Zone 1. Considering that the DVC influences voltages in the vicinity of its placement node, we identified the node with the highest voltage variations within the targeted zone. For the given sample feeder, candidate nodes were selected by evaluating the voltage variation profiles. The dispatching scheme uses a binary search algorithm \cite{lin2019binary} to determine the appropriate VAR injection/absorption required by the DVC on a per-phase basis. The following straightforward search procedure for candidate nodes determines which node yields optimal DVC performance:
\begin{enumerate}
    \item Place the DVC at a candidate node.
    \item Perform time series power flow simulation on the feeder over the sample days. Time resolution is 1 minute. The DVC is dispatched at every time step of the simulation by using the optimal DVC dispatch scheme introduced in Section II. B.
    \item Repeat the process by moving DVC to a new candidate bus.
\end{enumerate}

\section{Supervisory Dispatch for DVC}
\subsection{Optimal Q-V Trajectories}
\begin{figure}[t]
\centerline{\includegraphics[width=\columnwidth, height=0.19\textheight]{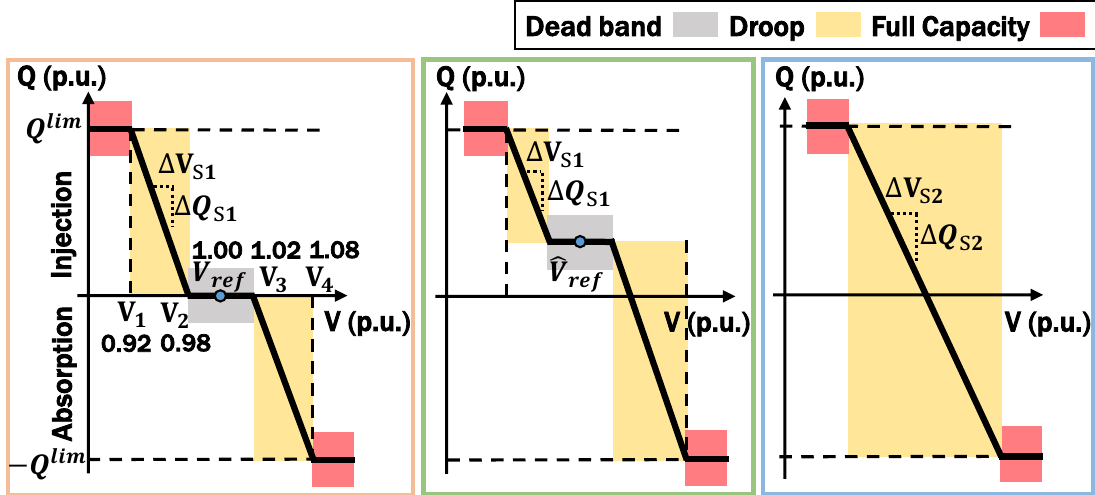}} \vspace{-.2in}
  \subfloat[\label{5a}]{\hspace{.40\columnwidth}}
  \subfloat[\label{5b}]{\hspace{.26\columnwidth}}
  \subfloat[\label{5c}]{\hspace{.35\columnwidth}}
  \vspace{-0.05in}
\caption{\textbf{Volt/VAR Curves (VV-Cs) for (a) Standard \cite{photovoltaics2018ieee}, (b) Shifted, and (c) Fitted.}}
\label{fig5}
\vspace{-0.2in}
\end{figure}

Figure~\ref{fig13} shows the optimal Q-V trajectories obtained by using the proposed optimal dispatch scheme on the sample system. The figure clearly illustrates that these optimal Q-V trajectories can be quite different than the VV-C proposed in IEEE Std. 1547-2018 \cite{photovoltaics2018ieee} for local control. The standard VV-C as shown in Fig.~\ref{fig5}(a), is a piecewise linear curve with negative slope. As formulated in \eqref{eq9}, when the voltage exceeds an upper limit (i.e., $V^\mathrm{upper}$), the DVC absorbs the reactive power to prevent further voltage rise. On the other hand, when the drops below a specific threshold (i.e., $V^\mathrm{lower}$), the DVC injects the reactive power to increase the voltage. 

\vspace{-0.1in}
\begin{equation}  \label{eq9} 
{Q_{inj}} = \begin{cases}
             Q^{lim},&{V_{dvc}(t) \leq V_{1}} \\
            -m_{1}(V_{2}-V_{dvc}(t)),&{V_{1} < V_{dvc}(t) < V_{2}} \\ 
             0,&{V_{2} \leq V_{dvc}(t) \leq V_{3}} \\ 
             m_{2}(V_{dvc}(t)-V_{3}),&{V_{3} < V_{dvc}(t) < V_{4}} \\  
            -Q^{lim},&{V_{4} \leq V_{dvc}(t)}
            \end{cases}
\end{equation}

\subsection{Supervisory Dispatch for DVC}
In Section~II.B, we considered the DVC as a dispatchable VAR source and employed an optimization-based dispatching scheme to continuously optimize its performance in terms of minimizing voltage variations. However, this approach faces a significant challenge due to the frequent dispatch signals required, which may not be practical in distribution systems with limited communication infrastructure \cite{neal2011advanced, manbachi2015real, muthukaruppan2020ami}. To overcome this challenge, a local control scheme, initially proposed for smart inverters and utilizing the VV-C specified in IEEE Std. 1547 (shown in Fig.~\ref{fig5}(a)), is currently utilized for the DVC. Nevertheless, to ensure the effectiveness of the DVC using this local control strategy, proper adjustment and setting of the VV-Cs are necessary. The optimal Q-V trajectories presented in the case study clearly illustrate the need for periodic adjustments. To address this issue, we investigated the problem and developed two supervisory control schemes that determine the optimal frequency of VV-C adjustments for the DVC to provide effective voltage support under varying operating conditions. These supervisory schemes monitor the performance of the DVC and make necessary adjustments to the VV-C, periodically sending the revised characteristics to the DVC. The proposed scheme involves two main steps: time segmentation and VV-C curve fitting based on the optimal Q-V profiles obtained for the respective time segment. The steps are outlined below. 

\subsubsection{Time Segmentation} 
\begin{figure}[t]
\centerline{\includegraphics[width=\linewidth, height=0.25\textheight]{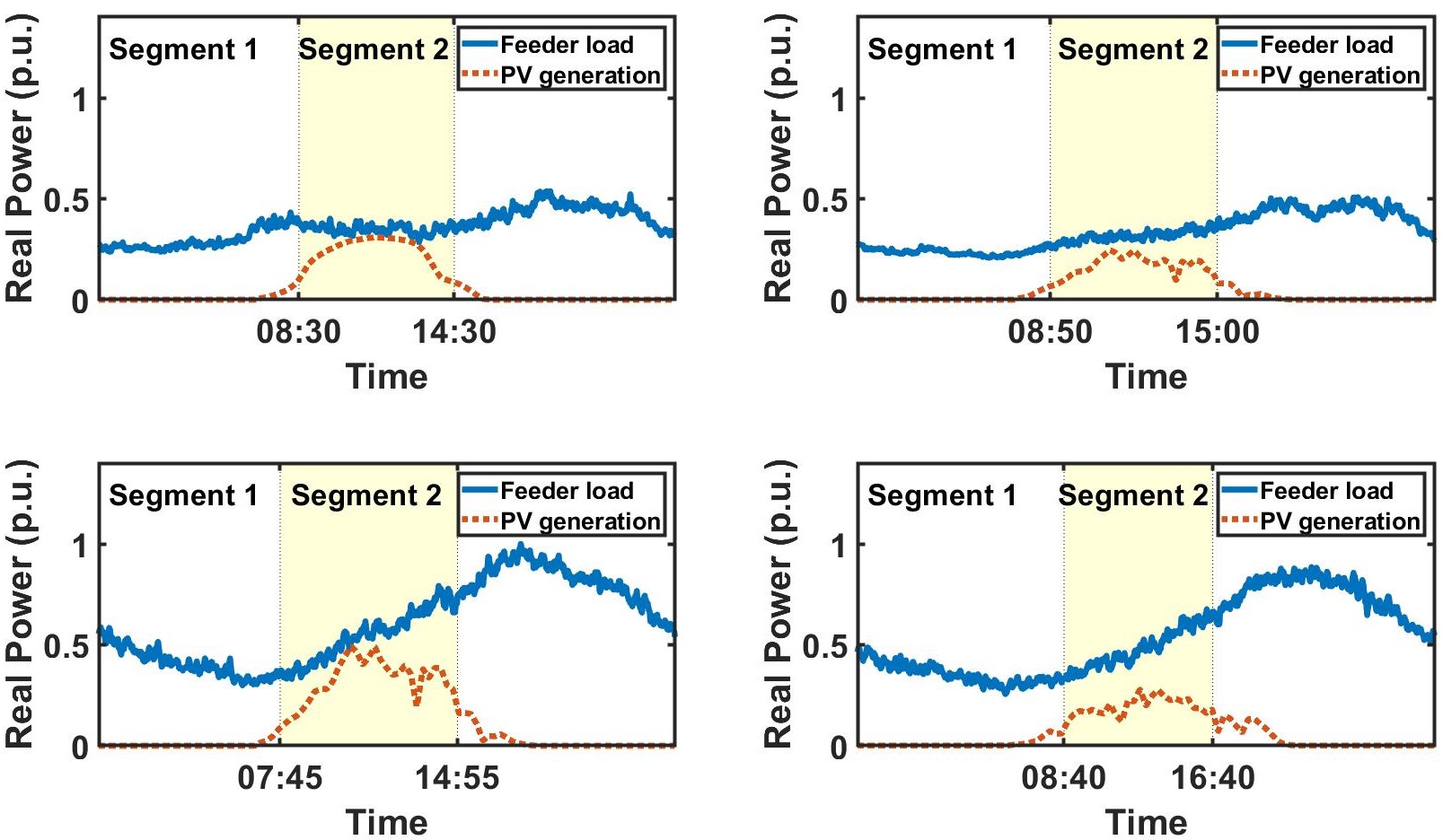}} \vspace{-1.42in}
  \subfloat[\label{6a}]{\hspace{.53\linewidth}}
  \subfloat[\label{6b}]{\hspace{.52\linewidth}} \vspace{+0.9in} \hfill
  \subfloat[\label{6c}]{\hspace{.53\linewidth}} 
  \subfloat[\label{6d}]{\hspace{.52\linewidth}} 
  \vspace{-0.05in}
\caption{\textbf{Time segmentation based on PV outputs, (a) Winter, (b) Spring, (c) Summer, and (d) Fall.}}
\label{fig6}
\vspace{-0.05in}
\end{figure}

The objective of time segmentation is to identify shorter time segments that allow for a good fit between the Q-V trajectories observed during these segments and the VV-C characteristics. Based on the results obtained from the optimal dispatch, it was observed that the voltage variations on the feeder are considerable during periods of highly variable PV output. Consequently, the Q dispatch of the DVC is adjusted accordingly to mitigate these variations. Conversely, when the PV output is low, the change in Q dispatch is not substantial. Therefore, the time segmentation is determined based on the PV output. In Fig.~\ref{fig6}, Segment 1 represents a period of low PV output when the PV generation is less than 25\% of the load, while Segment 2 corresponds to a period of high PV output (highlighted in yellow) when the PV generation exceeds 25\% of the load. By dividing the time into these distinct segments, we can better align the VV-C characteristics with the observed Q-V trajectories during different PV output conditions.

\vspace{-0.1in}
\subsubsection{Volt/VAR Curve (VV-C) Fitting} 
We propose two schemes for updating the VV-C for the DVC. The first scheme, called \textit{curve shifting}, involves shifting the midpoint of the standard VV-C (i.e., $\mathrm{V}_{ref}$) to align with the average Q-V point (i.e., $\mathrm{\hat{V}}_{ref}$) obtained from the optimal Q-V trajectory. Only the $\mathrm{V}_{ref}$ value is adjusted while maintaining the slope of the existing curve. In the second approach, called \textit{fitted VV-C}, we use linear regression \cite{weisberg2005applied} to determine the slope (i.e., $\Delta Q_{S2} / \Delta V_{S2}$) that best fits the VV-C to closely match the optimal Q-V trajectory. The curve settings are provided in Table~\ref{tab2}.

\begin{table}[t]
	\begin{center}
    \caption{\textbf{Volt/VAR Curve (VV-C) Settings.}}
    \label{tab2}
    \resizebox{\linewidth}{!}{%
    \begin{tabular}{ccc}
    \toprule
    {\textbf{VV-Cs}} & {\textbf{Slope}} & {\textbf{Dead band}} \\ 
	\midrule
	\textbf{Standard} & $\Delta Q_{S1} / \Delta V_{S1} = -Q^{lim} / (V_{2}-V_{1})$ & $[V_{2}\ V_{3}] = [0.98\ 1.02]$ \\
	\textbf{Shifted}  & $\Delta Q_{S1} / \Delta V_{S1}$ & $[\hat{V}_{ref}-0.02\quad \hat{V}_{ref}+0.02]$ \\
	\textbf{Fitted}   & $\Delta Q_{S2} / \Delta V_{S2}$ & - \\
   \bottomrule 
   \end{tabular}}
   \end{center}
\vspace{-0.3in}
\end{table}

The next step is to determine the frequency at which the VV-C should be updated to ensure effective voltage support under varying operating conditions. As illustrated in Fig.~\ref{fig6}, Segment 1 experiences low PV output, and thus the IEEE Std. 1547 VV-C is adopted. In Segment 2, with significant PV output, the VV-C is updated using the optimal dispatch results obtained for this segment. It is worth noting that the ideal approach would involve utilizing the optimal Q dispatch and voltage for the subsequent interval. Established methods, such as statistical or neural network-based approaches \cite{li2021meta, li2023tcn}, can be employed for short-term load and solar PV forecasting to facilitate this process. However, this is not the focus of this paper, therefore the simplest prediction available is to assume that we already know the predictions for the next interval.

\begin{figure}[t]
\centerline{\includegraphics[width=0.95\columnwidth, height=0.18\textheight]{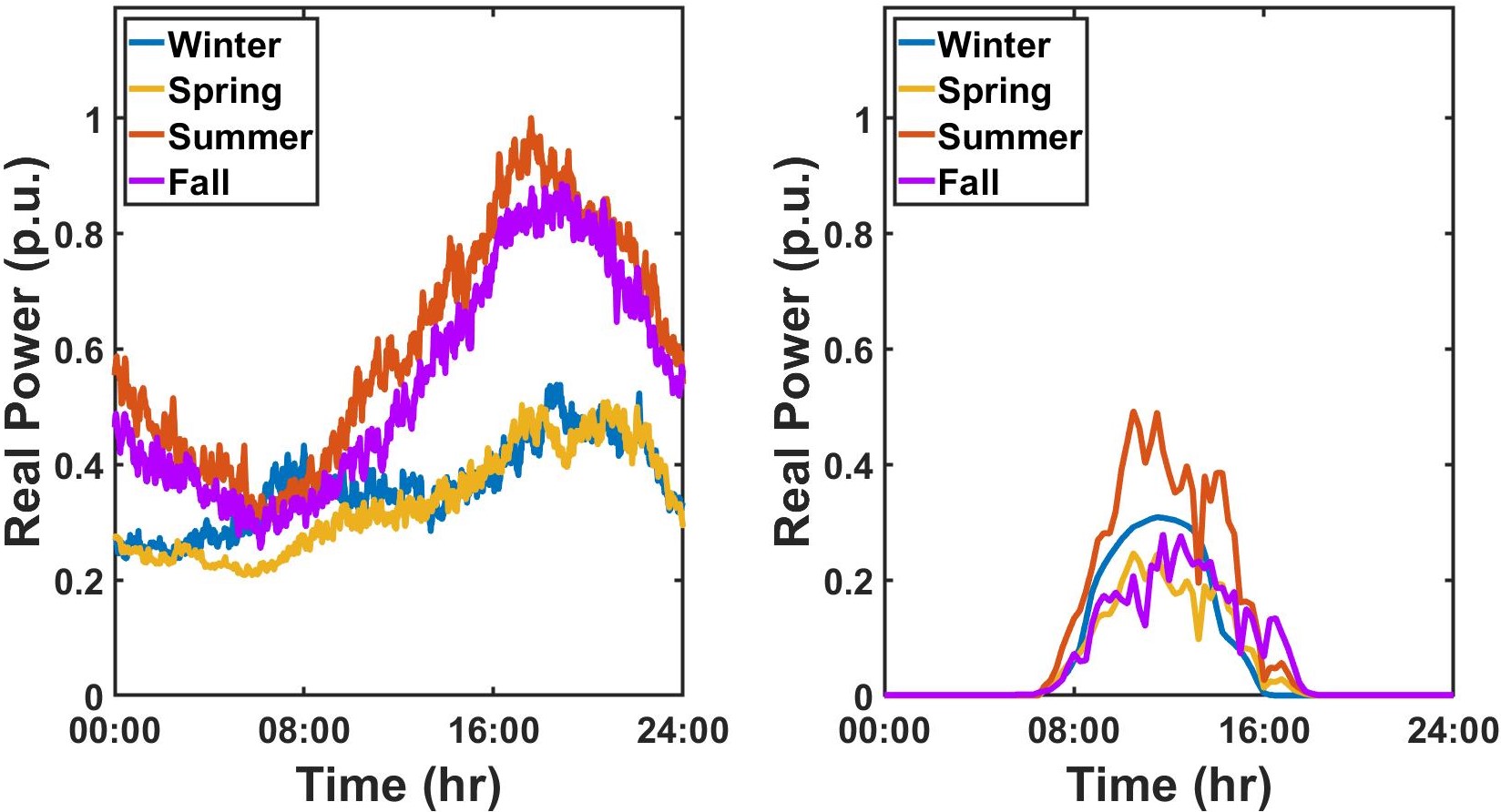}} \vspace{-.2in}
  \subfloat[\label{7a}]{\hspace{.54\columnwidth}}
  \subfloat[\label{7b}]{\hspace{.45\columnwidth}}
\vspace{-0.05in}
\caption{\textbf{Real power of (a) Feeder load, (b) PV outputs.}}
\vspace{-0.1in}
\label{fig7}
\end{figure}

\section{Case Study}
The IEEE 123 node test system shown in Fig.~\ref{fig3} is used to test and demonstrate the effectiveness of the proposed DVC optimal dispatching scheme in unbalanced scenario. This feeder is rated at 4.16kV and the substation transformer is equipped with a load tap changer (LTC). Additionally, there are 6 single-phase load voltage regulators (LVRs) for voltage regulation. To simulate high PV penetration on the feeder, five 1 MW PV farms are placed at nodes 18, 47, 54, 76, and 101, and a 1 MVAR 3-phase DVC is also considered. OpenDSS is used to do the time series power flow simulations and the DVC is modelled as three single-phase impedance banks with independent control on each phase. The load and PV profiles utilized in this study are obtained from two different data sources. The 1-minute smart meter data sets are sourced from the Pecan Street data repository \cite{street2019dataport}, while the 1-minute PV data sets are collected from Duke Energy in North Carolina. The ZIP load model in OpenDSS is implemented with model 8 by setting ZIPV = [0.24, 0.36, 0.40, 0.24, 0.36, 0.40, 0.80]. Figure~\ref{fig7} presents the normalized load and PV profiles for four selected sample days. 

\subsection{DVC placement}
The proposed approach is applied to determine the optimal location for the DVC in the system. Firstly, the node voltage variations in Zone 1, where the DVC is intended to be placed, are obtained without the DVC (i.e., base case). The node voltage profiles obtained are depicted in Fig.~\ref{fig4}. Based on these profiles, three candidate nodes (i.e., nodes 7, 8, and 13) are selected as they have the largest voltage variations. Subsequently, the DVC is positioned at these candidate locations, and the optimal dispatch is used to evaluate the effectiveness of the DVC in mitigating voltage variations on the feeder. To determine the DVC placement, only the voltage variation (i.e., $f^{\mu}$ with $w_\mathrm{\mu}=1$ and $f^{\theta}$ with $w_\mathrm{\theta}=0$) is considered as the main objective for the DVC dispatch in \eqref{eq3}. Table~\ref{tab3} presents the performance metrics obtained for these three scenarios. At each time step, the optimal Q dispatch of the DVC is determined to minimize voltage variations while monitoring the voltage levels of all nodes in the test system. The total number of voltage points (T) monitored during the scheduling period is 1,549,440.

\begin{table}[b]
	\begin{center}
        \vspace{-0.2in}
		\caption{\textbf{Voltage Violations of 3 Candidate Nodes.}}   \vspace{-0.1in}
		\label{tab3}
		\centerline{\includegraphics[width=\linewidth]{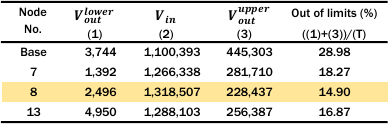}}
	\end{center}
 \vspace{-0.2in}
\end{table}

\begin{figure}[t]
\vspace{-0.1in}
\centerline{\includegraphics[width=\linewidth]{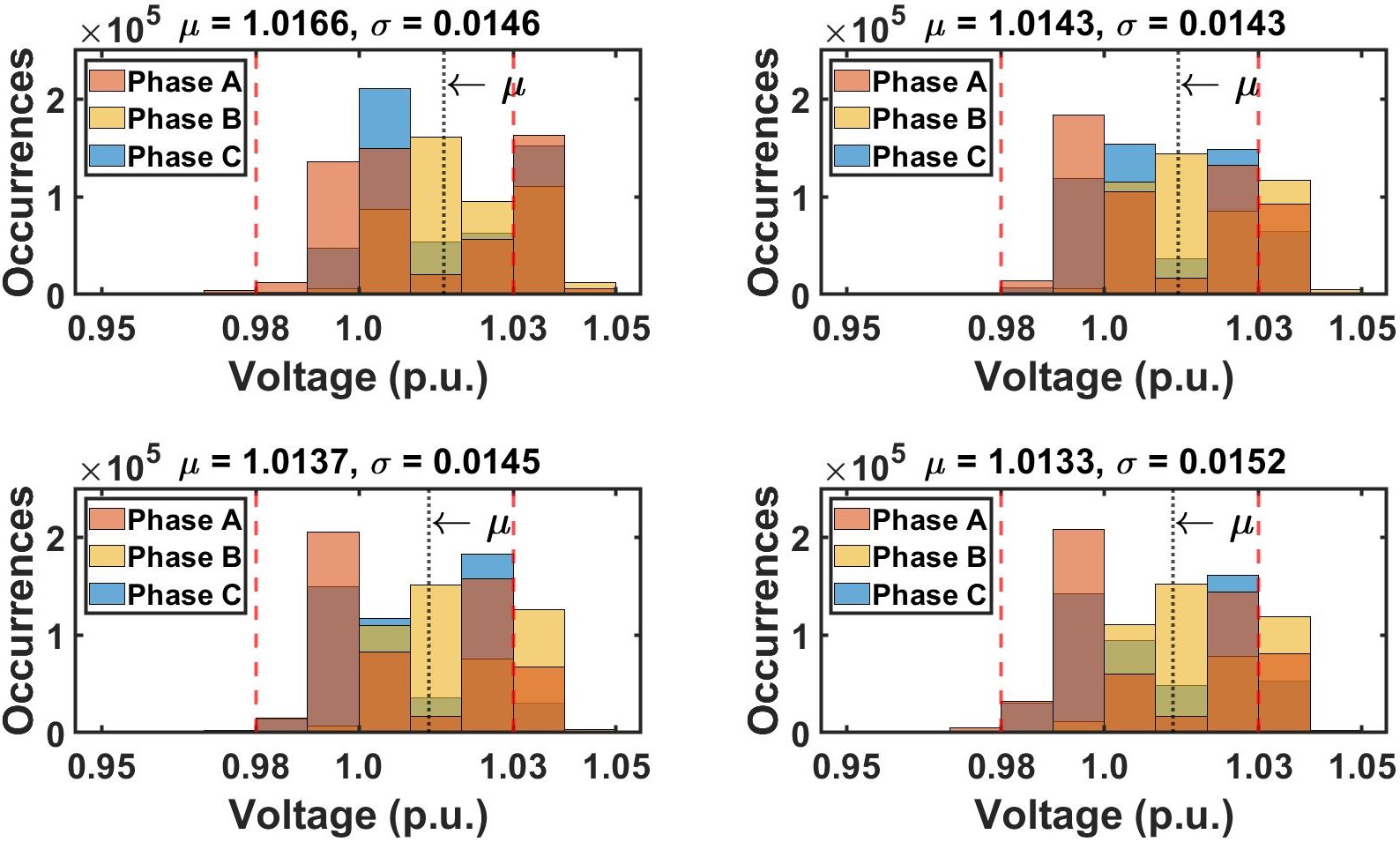}} \vspace{-1.27in}
  \subfloat[\label{8a}]{\hspace{.53\linewidth}}
  \subfloat[\label{8b}]{\hspace{.54\linewidth}} \vspace{+0.73in} \hfill
  \subfloat[\label{8c}]{\hspace{.53\linewidth}} 
  \subfloat[\label{8d}]{\hspace{.54\linewidth}} 
  \vspace{-0.05in}
\caption{\textbf{Voltage distribution by DVC placement, (a) no DVC, (b) node~7, (c) node~8, and (d) node~13.}}
\vspace{-.05in}
\label{fig8}
\end{figure}

\begin{figure}[ht]
\centerline{\includegraphics[width=\columnwidth, height=0.2\textheight]{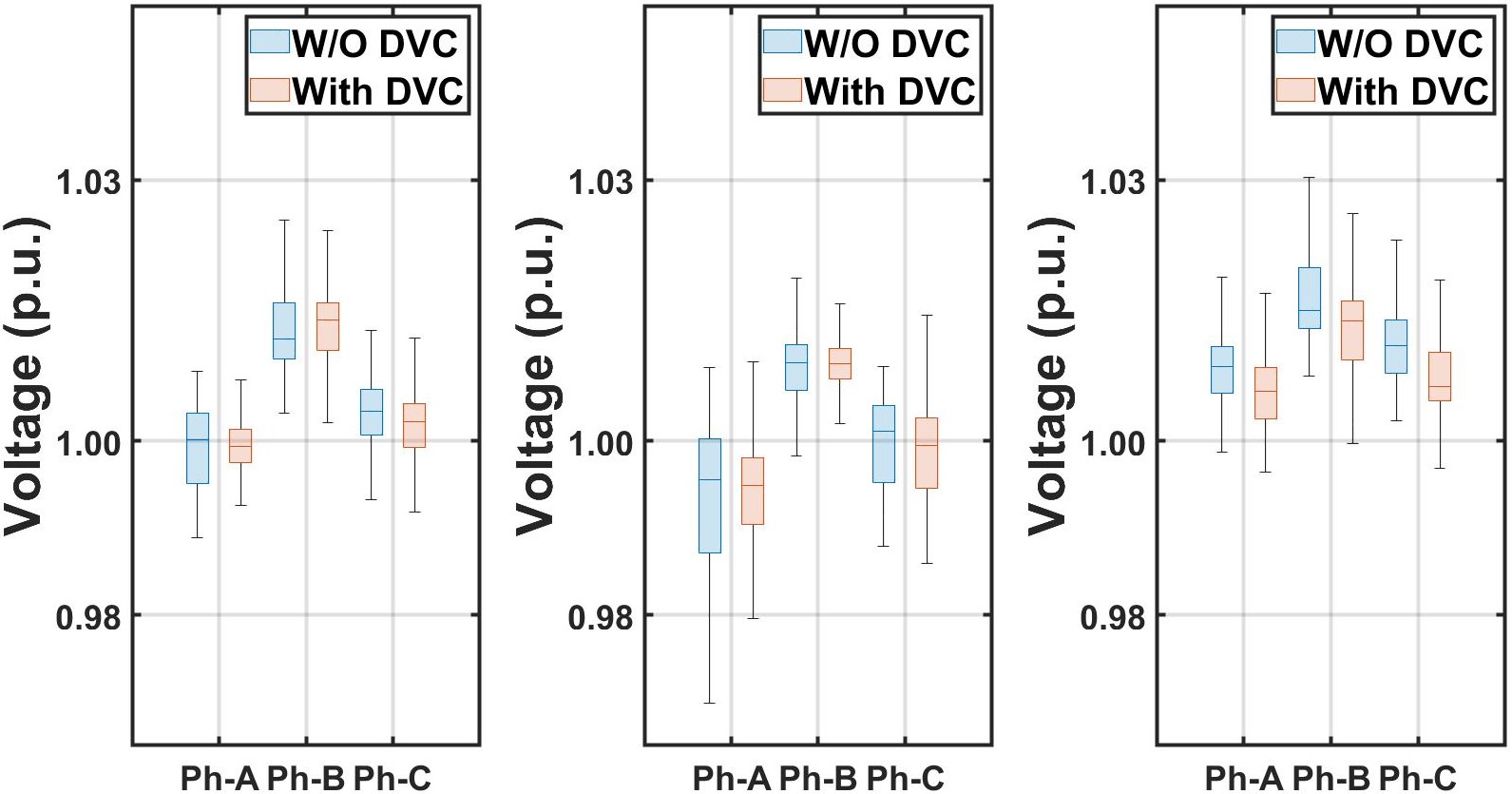}} \vspace{-.2in}
  \subfloat[\label{9a}]{\hspace{.40\columnwidth}}
  \subfloat[\label{9b}]{\hspace{.28\columnwidth}}
  \subfloat[\label{9c}]{\hspace{.40\columnwidth}}
  \vspace{-0.05in}
\caption{\textbf{Distributions of voltage variations without and with DVC at node~8 for (a) node~29, (b) node~66, and (c) node~8.}}
\vspace{-.2in}
\label{fig9}
\end{figure}

Figure~\ref{fig8} shows the node voltage histograms for the three scenarios, revealing the impact of placing the DVC at these locations on reducing voltage variations among the feeder nodes. The results demonstrate a notable improvement, as a significant portion of node voltages now fall within the desired voltage band. Specifically, the percentage of node voltages outside the band decreases from 28.98\% in the base case to 14.90\% when the DVC is placed at node 8. Moreover, the voltage variation statistics show slight variations across the different phases of the circuit. In all scenarios, the average ($\mu$) voltage decreases compared to the base case, while the standard deviation ($\sigma$) shows varying changes. Ultimately, after considering the performance metrics, node 8 is chosen as the optimal location since it yields the most favorable statistics for both the lower and upper voltage bands.

Figure~\ref{fig9} provides an evaluation of the performance of the DVC by examining voltage variations at three selected nodes (29, 66, and 8) in Zone 1, both with and without the presence of the DVC at node 8. Node 8 represents the location where the DVC is placed, while nodes 29 and 66 are the farthest nodes connected to the mainline within the same zone. As depicted in the Fig.~\ref{fig9}, the DVC demonstrates a noticeable reduction in the occurrences of low voltages (< 0.98) and high voltages (> 1.03). However, note that the impact of the DVC on nodes 29 and 66 is minimal, with only slight changes observed. Conversely, the DVC significantly diminishes voltage variations at the node to which it is connected. This observation suggests that the DVC is particularly effective in reducing voltages at the bus it is connected and neighboring buses.

\subsection{Optimal Dispatch}
The placement of the DVC at node 8 (i.e., Case 1) introduces an undesirable effect, leading to an increase in LTC and LVR operations compared to the base case (i.e., Case 0), as shown in Table~\ref{tab5} and Figure~\ref{fig11}. The results highlight a significant increase in tap operations. This issue emphasizes the need for an optimal DVC dispatching approach that considers two objectives: $f^\mathrm{\mu}$, the primary objective aimed at minimizing voltage variations, and $f^\mathrm{\theta}$, the secondary objective aimed at limiting LVR tap changes. Since the number of tap operations is numerically a large value compared to $f^\mathrm{\mu}$, we tried with two different weights for $f^\mathrm{\theta}$: 1 and 0.1. To determine the most suitable option among these alternatives, we simulated the following four cases:
\begin{itemize}
    \item Case~0 (Base Case): This is the base case which corresponds to the system without the DVC.
    \item Case~1: This case only considers the voltage variation ($f^\mathrm{\mu}$) as the main objective for the DVC dispatch. The dispatching scheme is employed to determine the appropriate VAR injection/absorption required for the DVC to minimize voltage variations.
    \item Case~2: In this case, the objective for the DVC dispatch combines both the voltage variation metric $f^\mathrm{\mu}$ with $w_\mathrm{\mu}=1$ and tap changes metric $f^\mathrm{\theta}$ with $w_\mathrm{\theta}=1$. 
    \item Case~3: This case is the same as Case 2 but the weight for the LVR tap metric $f^\mathrm{\theta}$ is reduced to $w_\mathrm{\theta}=0.1$.
\end{itemize}

\begin{table}[b]
\vspace{-0.1in}
	\begin{center}
		\caption{\textbf{Voltage Violations by Case.}}    \vspace{-0.1in}
		\label{tab4}
		\centerline{\includegraphics[width=\linewidth]{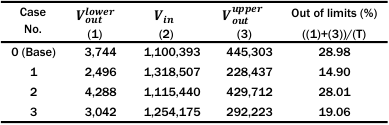}}
	\end{center}
 \vspace{-0.1in}
\end{table}

\begin{table}[b]
\vspace{-0.05in}
	\begin{center}
		\caption{\textbf{LTC and LVR Tap Changes by Case.}}   \vspace{-0.1in}
		\label{tab5}
		\centerline{\includegraphics[width=\linewidth]{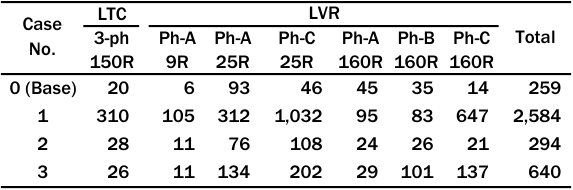}}
	\end{center}
 \vspace{-0.1in}
\end{table}

\begin{figure}[t]
\centerline{\includegraphics[width=\linewidth]{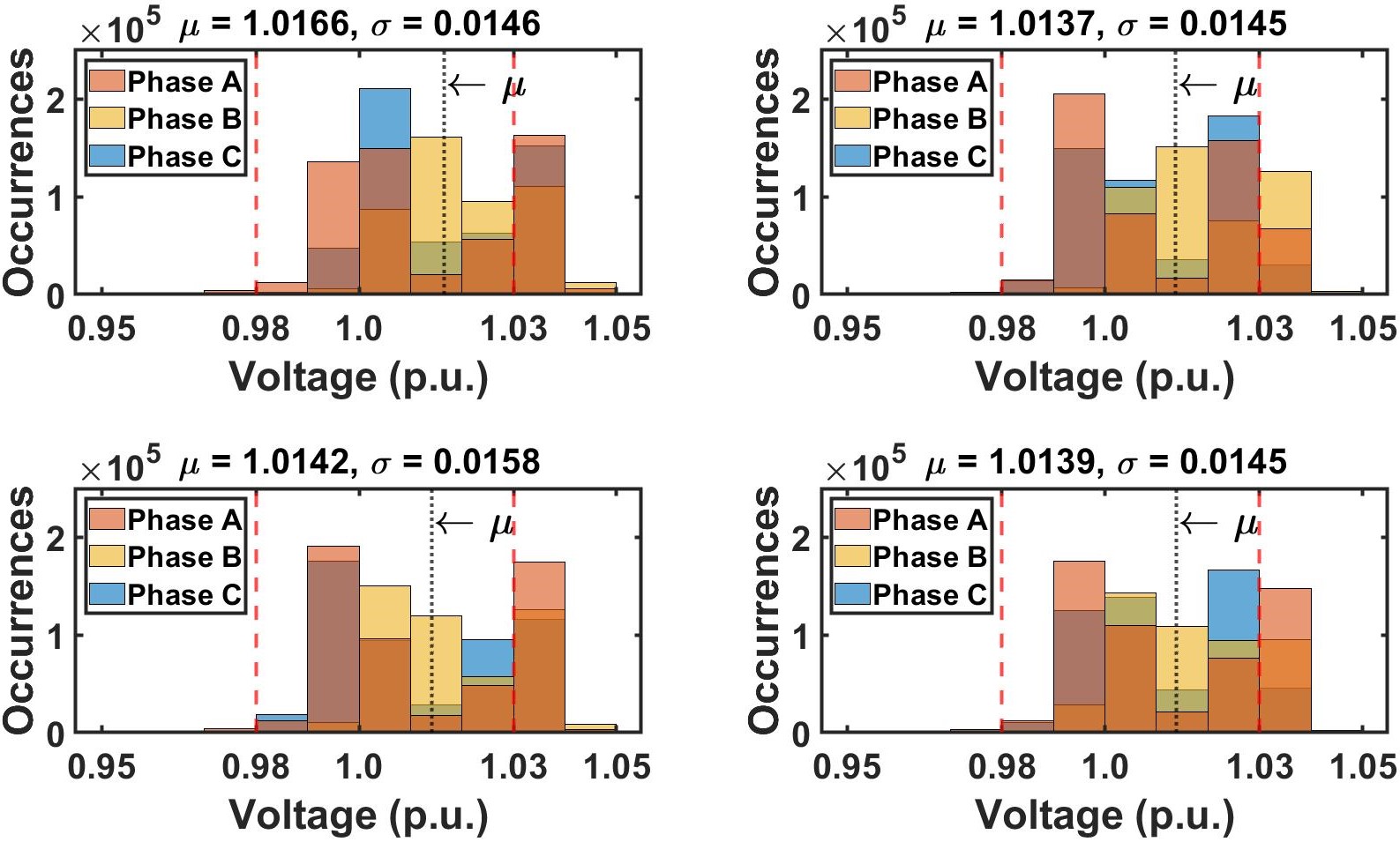}} \vspace{-1.29in}
  \subfloat[\label{10a}]{\hspace{.53\linewidth}}
  \subfloat[\label{10b}]{\hspace{.54\linewidth}} \vspace{+0.75in} \hfill
  \subfloat[\label{10c}]{\hspace{.53\linewidth}} 
  \subfloat[\label{10d}]{\hspace{.54\linewidth}} 
  \vspace{-0.05in}
\caption{\textbf{Voltage distribution by case, (a) Case~0, (b) Case~1, (c) Case~2, and (d) Case~3.}}
\vspace{-0.1in}
\label{fig10}
\end{figure}

\begin{figure}[t]
\centerline{\includegraphics[width=\linewidth]{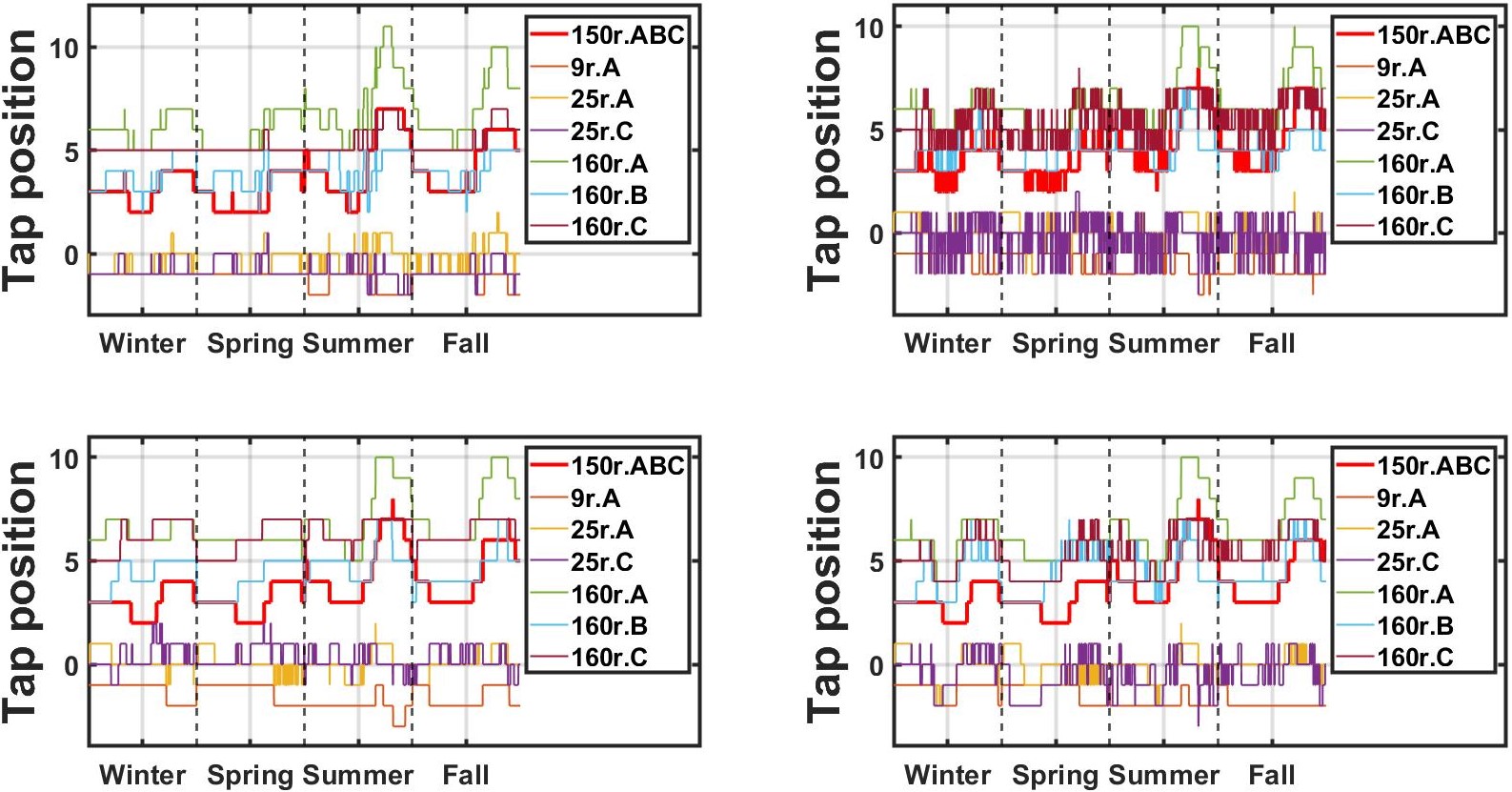}} \vspace{-1.17in}
  \subfloat[\label{11a}]{\hspace{.53\linewidth}}
  \subfloat[\label{11b}]{\hspace{.54\linewidth}} \vspace{+0.63in} \hfill
  \subfloat[\label{11c}]{\hspace{.53\linewidth}} 
  \subfloat[\label{11d}]{\hspace{.54\linewidth}} 
  \vspace{-0.05in}
\caption{\textbf{LTC and LVR tap changes by case, (a) Case~0, (b) Case~1, (c) Case~2, and (d) Case~3.}}
\vspace{-0.25in}
\label{fig11}
\end{figure}

\begin{figure}[b]
\vspace{-0.25in}
\centerline{\includegraphics[width=\linewidth]{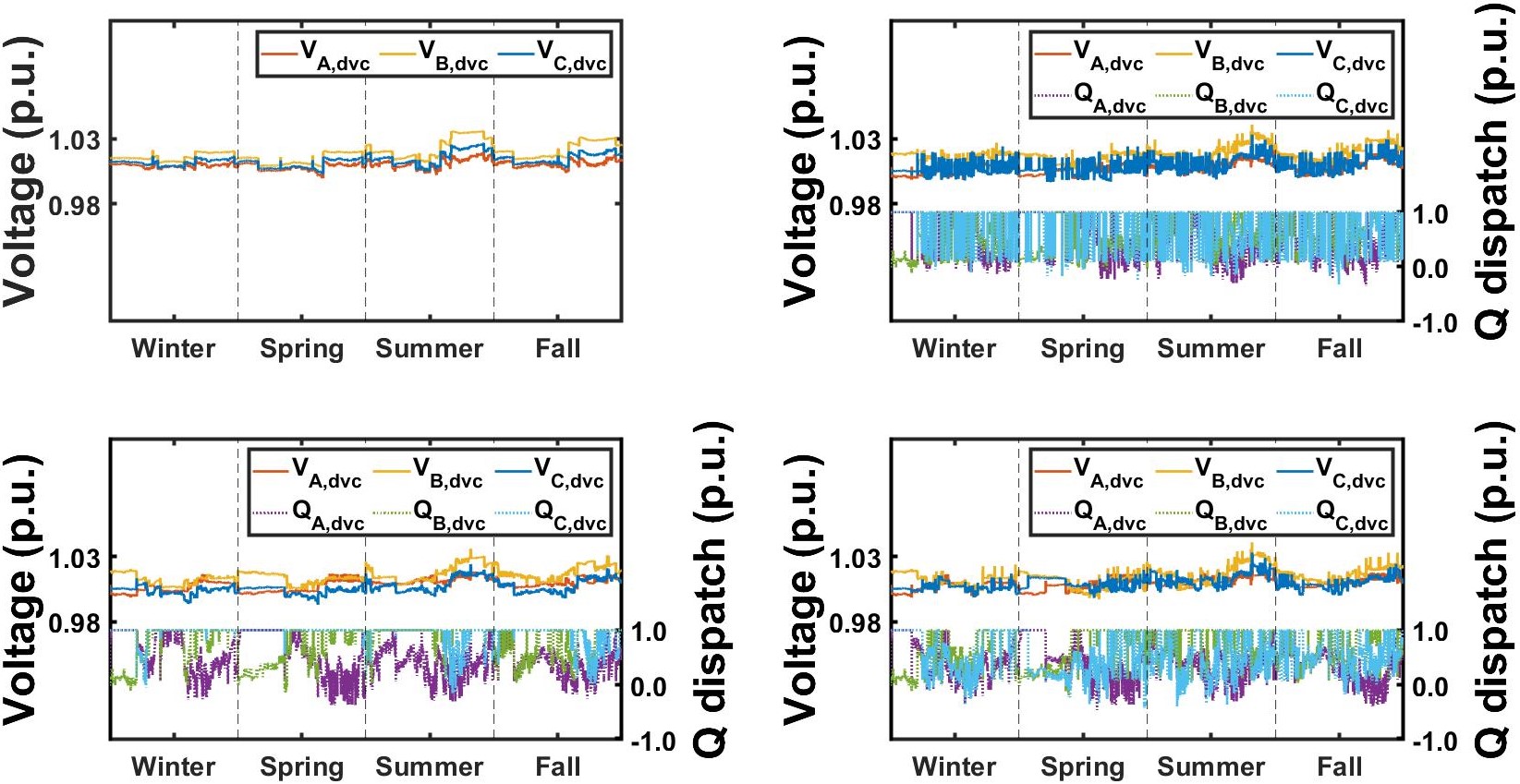}} \vspace{-1.14in}
  \subfloat[\label{12a}]{\hspace{.48\linewidth}}
  \subfloat[\label{12b}]{\hspace{.53\linewidth}} \vspace{+0.61in} \hfill
  \subfloat[\label{12c}]{\hspace{.48\linewidth}} 
  \subfloat[\label{12d}]{\hspace{.53\linewidth}} 
  \vspace{-0.05in}
\caption{\textbf{Voltage and Q dispatch of DVC for (a) Base, (b) Case~1, (c) Case~2, and (d) Case~3.}}
\label{fig12}
\end{figure}

Simulation results for these four cases are summarized in Tables~\ref{tab4} and \ref{tab5}. The key observations are summarized below:
\begin{itemize}
    \item Compared to Case 0 (base case), Cases 1, 2, and 3 all reduce node voltage variations, as indicated by the performance statistics presented in Table~\ref{tab4}. Figure~\ref{fig10} shows the voltage distribution for the four cases, highlighting how the voltages are shifted closer to the desired voltage band. 
    \item Figure~\ref{fig11} compares the number of LTC and LVR operations across different cases. The results demonstrate that focusing only on voltage variation in the dispatch (Case~1) leads to an increase in LVR operations. However, Case~3, which incorporates the revised objective, provides a good compromise by reducing LVR operations compared to Case~1, without degrading the voltage variation performance of the DVC.
    \item Figure~\ref{fig12} shows the optimal Q dispatch results and voltage at node 8 for each case. The results reveal that the DVC primarily injects reactive power (kVAR) throughout the duration. Furthermore, the terminal voltage at node 8 consistently remains in close proximity to the upper voltage band limit of 1.03 pu.
\end{itemize}

\subsection{Selecting Weights for Dispatch with Combined Objectives}
Based on the aforementioned findings, it can be inferred that the voltage variation outcomes are influenced by the weight assigned to tap change metrics. Therefore, a sensitivity analysis is performed to assess the effects of varying tap change weights on the results. The simulations are repeated using different weights of $w_\mathrm{\theta}$=\{0.01, 0.05, 0.1, 0.5\}.

\begin{table}[b]
\vspace{-0.2in}
	\begin{center}
		\caption{\textbf{Voltage Violations by Tap Change Weight.}}   \vspace{-0.1in}
		\label{tab6}
		\centerline{\includegraphics[width=\linewidth]{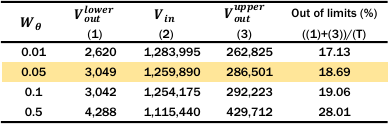}}
	\end{center}
 \vspace{-.1in}
\end{table}

\begin{table}[b]
\vspace{-0.05in}
	\begin{center}
		\caption{\textbf{LTC and LVR Tap Changes by Tap Change Weight.}} \vspace{-0.1in}
		\label{tab7}
		\centerline{\includegraphics[width=\linewidth]{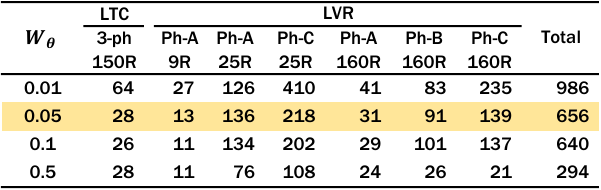}}
	\end{center}
 \vspace{-.1in}
\end{table}

\begin{figure}[t]
\vspace{-.05in}
\centerline{\includegraphics[width=\linewidth, height=0.25\textheight]{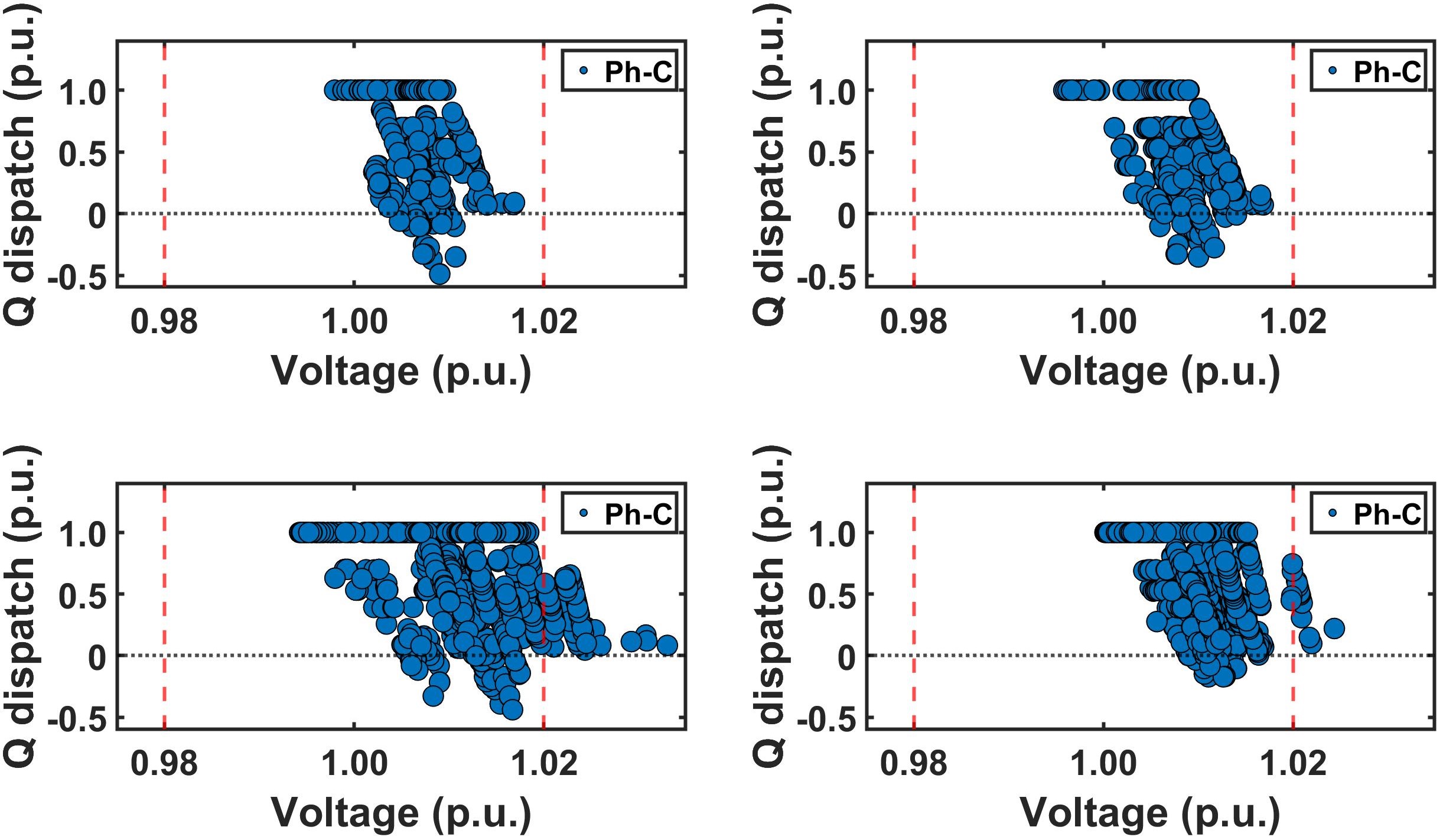}} \vspace{-1.4in}
  \subfloat[\label{13a}]{\hspace{.56\linewidth}}
  \subfloat[\label{13b}]{\hspace{.46\linewidth}} \vspace{+0.9in} \hfill
  \subfloat[\label{13c}]{\hspace{.56\linewidth}} 
  \subfloat[\label{13d}]{\hspace{.46\linewidth}} 
  \vspace{-0.05in}
\caption{\textbf{Optimal Q-V points of Phase~C for (a) Winter, (b) Spring, (c) Summer, and (d) Fall.}}
\vspace{-0.2in}
\label{fig13}
\end{figure}

\begin{figure*}[ht]
\vspace{-.1in}
    \subfloat[\textbf{Segment~2-1}]{%
        \includegraphics[width=.48\linewidth, height=0.15\textheight]{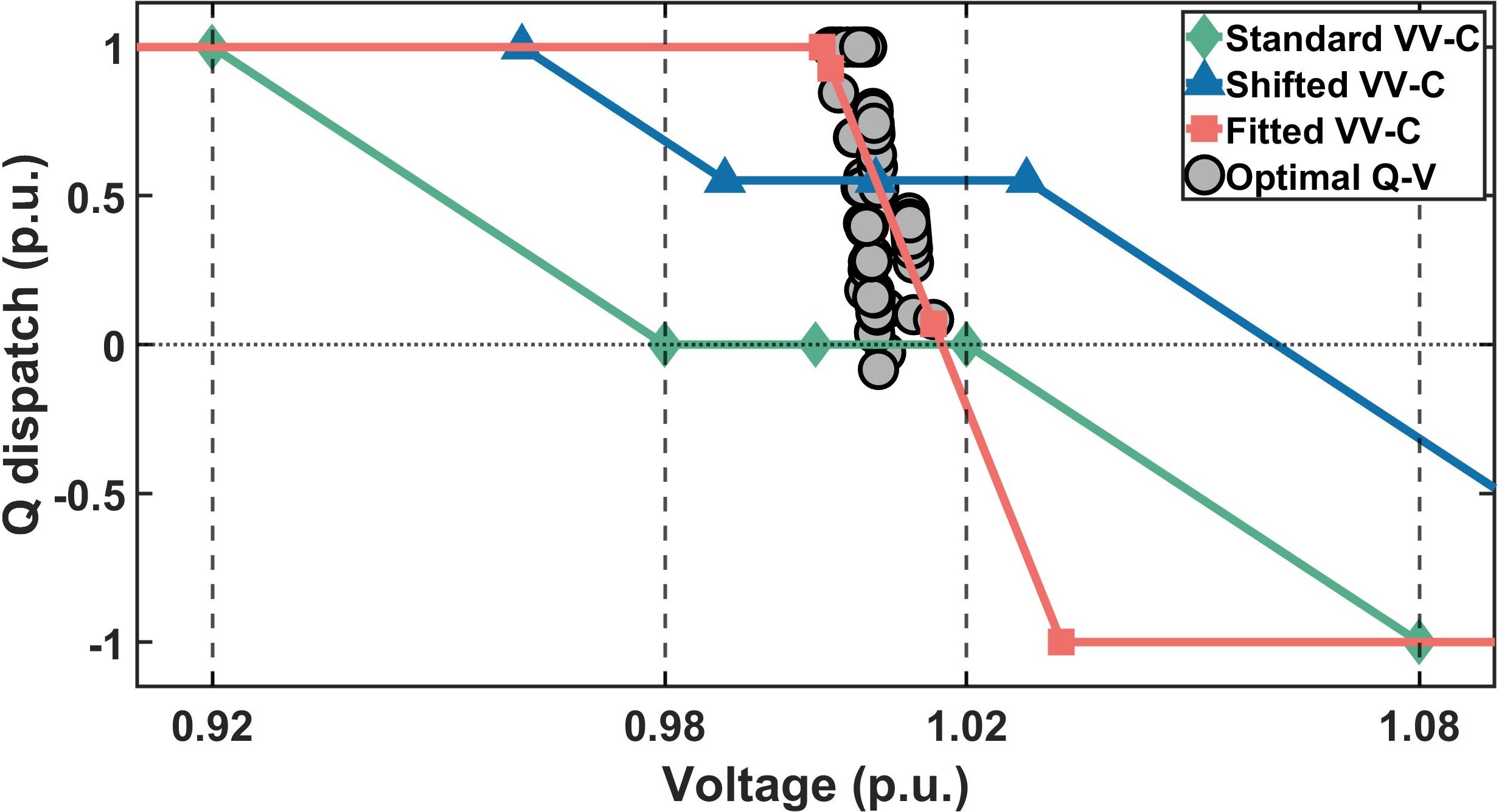}%
        \label{14a}%
    }\hfill
    \subfloat[\textbf{Segment~2-2}]{%
        \includegraphics[width=.48\linewidth, height=0.15\textheight]{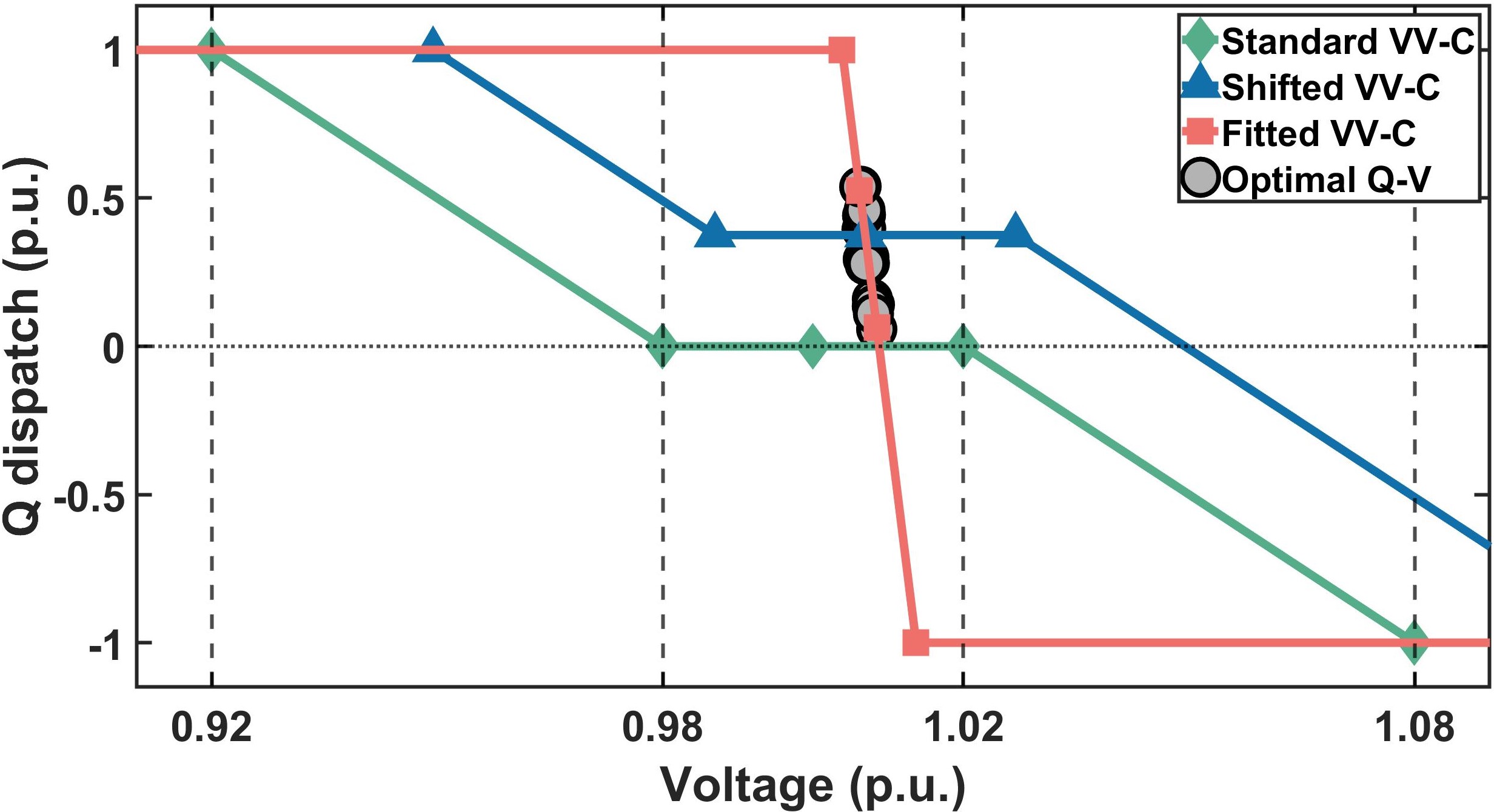}%
        \label{14b}%
    }\hfill \\
    \subfloat[\textbf{Segment~2-3}]{%
        \includegraphics[width=.48\linewidth, height=0.15\textheight]{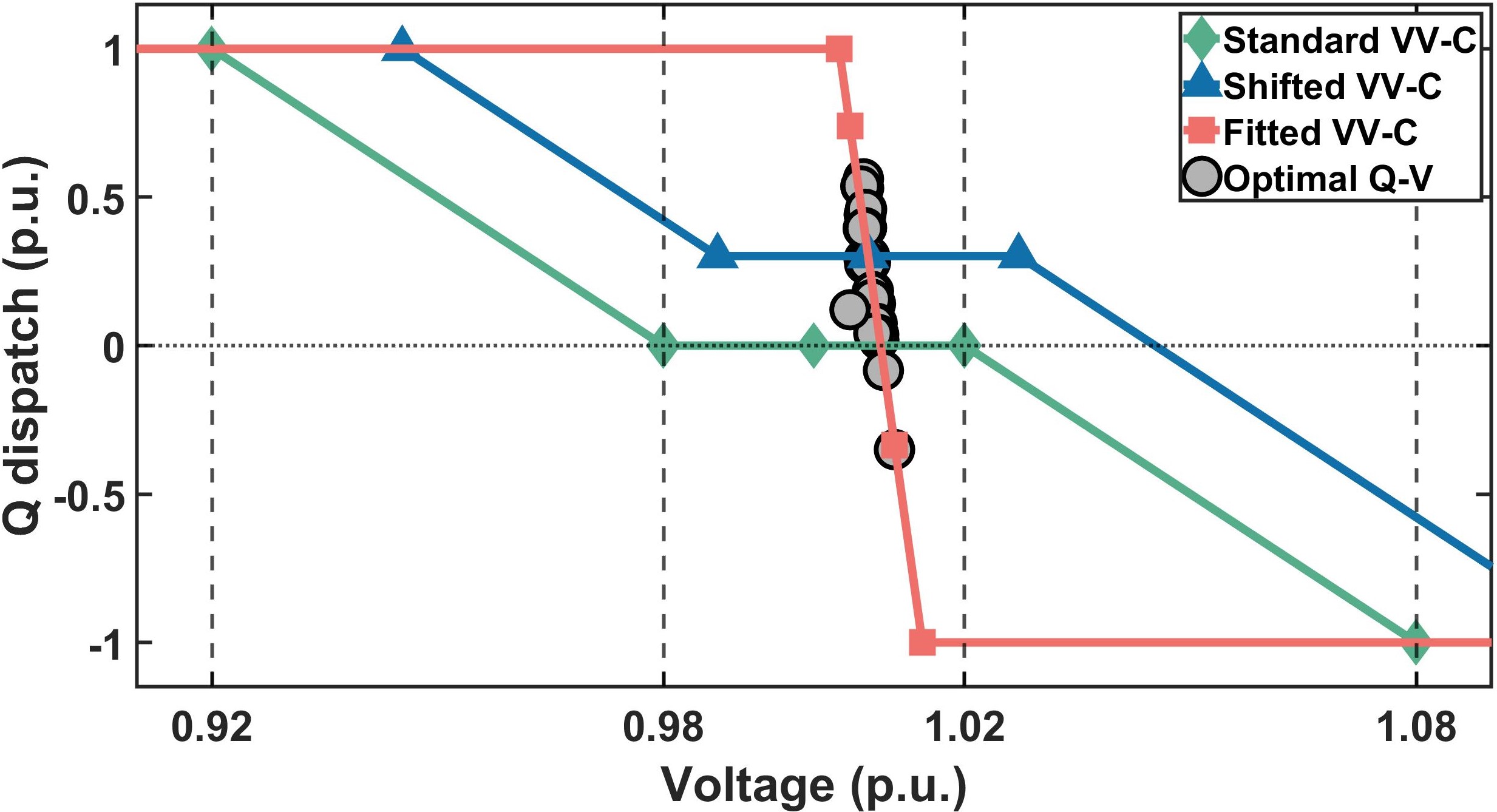}%
        \label{14c}%
    }\hfill
    \subfloat[\textbf{Segment~2-4}]{%
        \includegraphics[width=.48\linewidth, height=0.15\textheight]{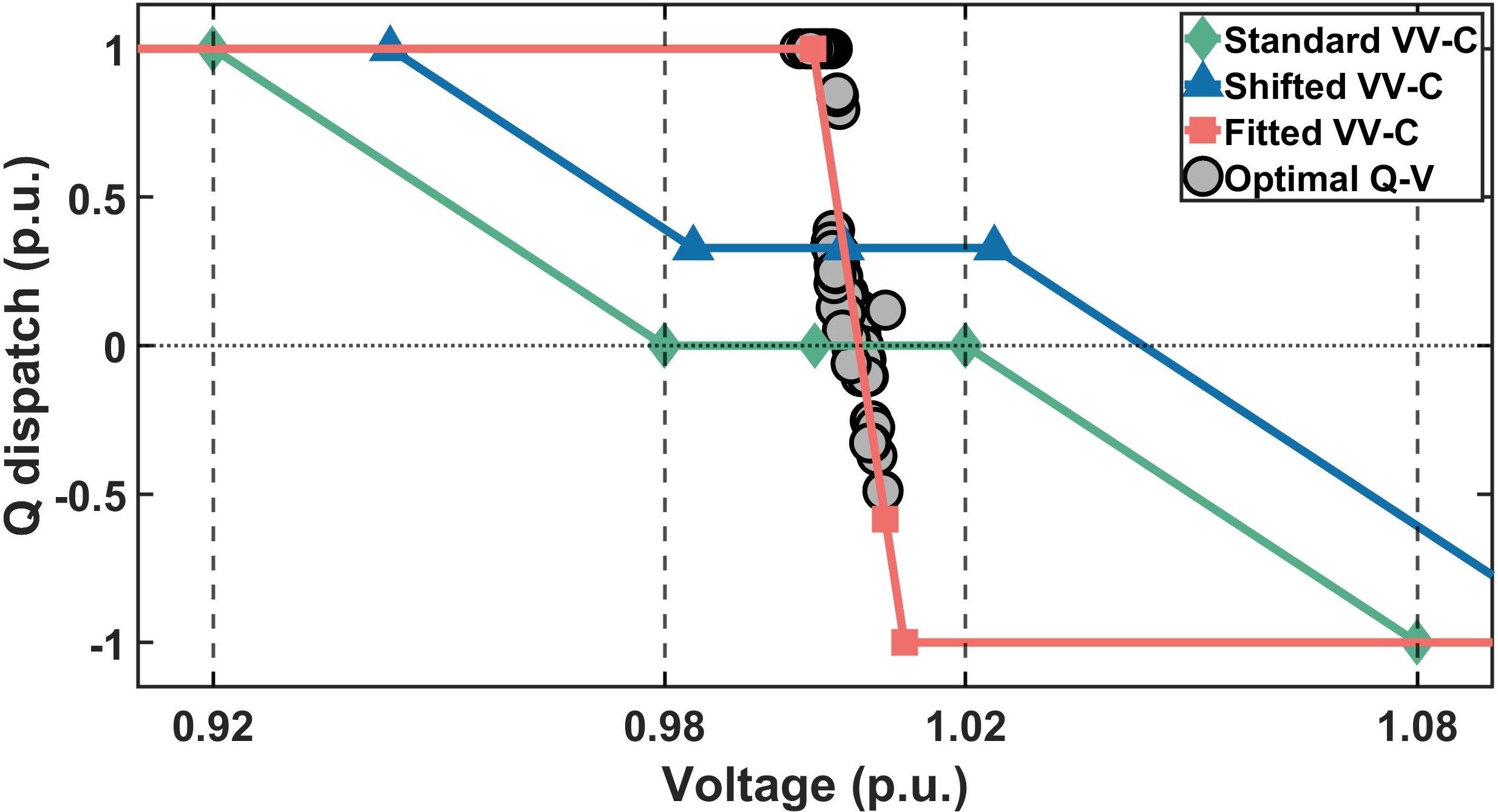}%
        \label{14d}%
    }
    \vspace{-0.05in}
    \caption{\textbf{Optimal Q dispatch of the DVC at Phase C in winter and local control schemes for (a) 08:00-10:00, (b) 10:00-12:00, (c) 12:00-14:00, and (d) 14:00-16:00.}} 
    \vspace{-.1in}
    \label{fig14}
\end{figure*}

\begin{figure}[ht]
\centerline{\includegraphics[width=0.95\columnwidth, height=0.16\textheight]{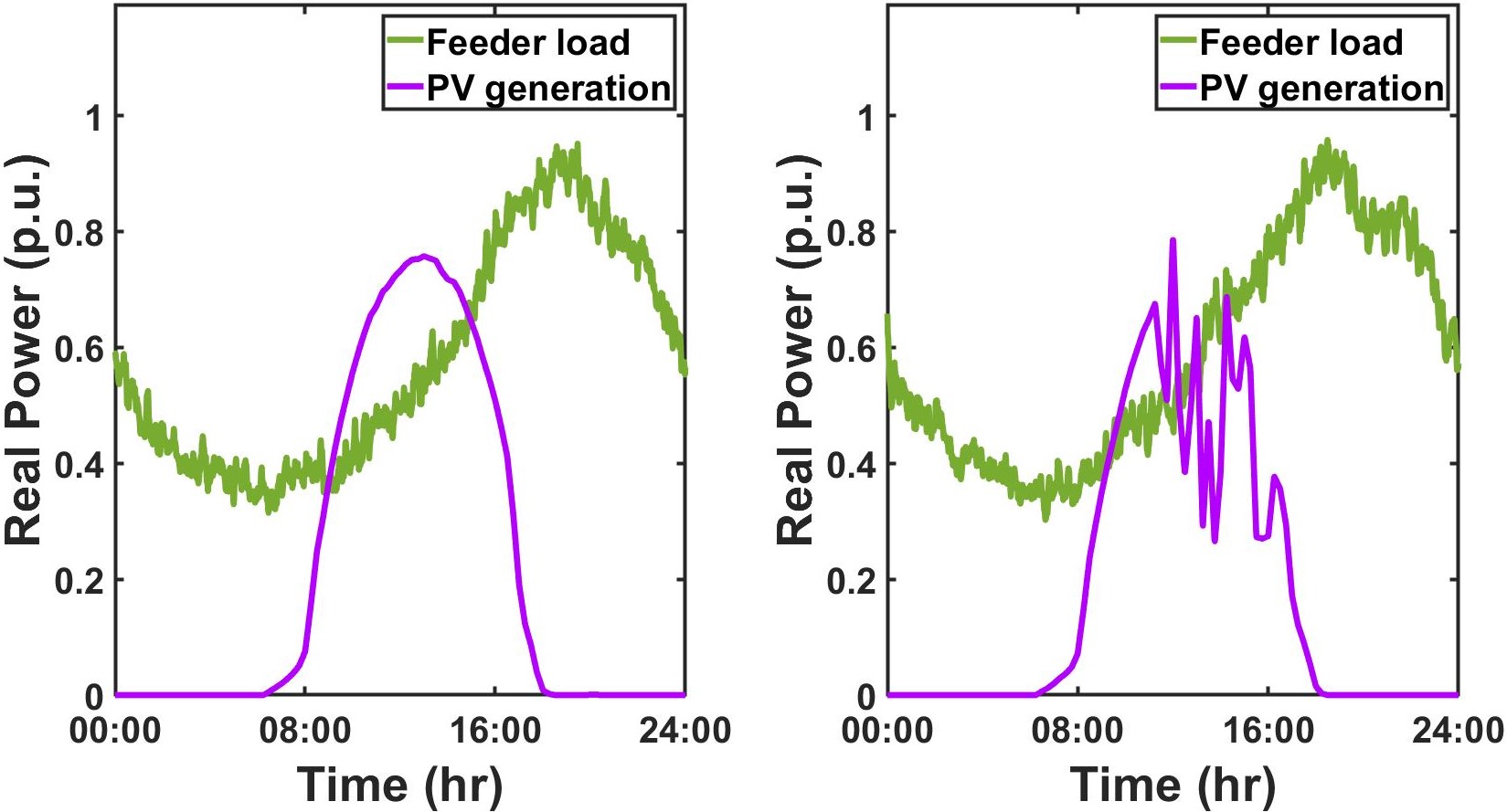}} \vspace{-.2in}
  \subfloat[\label{15a}]{\hspace{.55\linewidth}}
  \subfloat[\label{15b}]{\hspace{.45\linewidth}} 
  \vspace{-0.05in}
\caption{\textbf{Real power profile of load and PV for (a) Sunny day, (b) Cloudy day.}}
\vspace{-0.2in}
\label{fig15}
\end{figure}

\begin{table}[t]
\vspace{-0.1in}
	\begin{center}
		\caption{\textbf{Voltage Violation by Different VVC.}}    \vspace{-0.1in}
		\label{tab8}
		\centerline{\includegraphics[width=\linewidth]{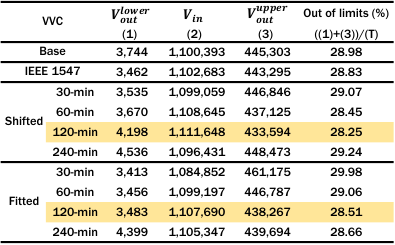}}
	\end{center}
\vspace{-0.1in}
\end{table}

\begin{table}[ht]
	\begin{center}
		\caption{\textbf{LTC and LVR Tap Changes by Different VVC.}}  \vspace{-0.1in}
		\label{tab9}
		\centerline{\includegraphics[width=\linewidth]{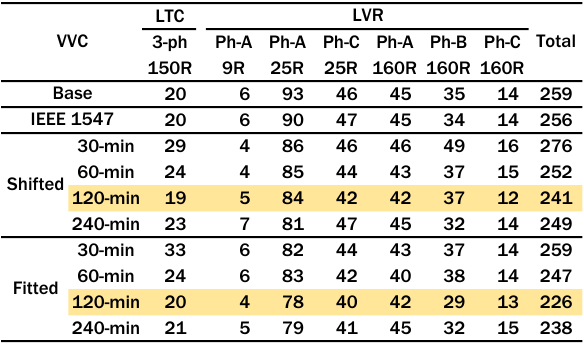}}
	\end{center}
\vspace{-0.2in}
\end{table}

The results presented in Tables~\ref{tab6} and \ref{tab7} demonstrate the importance of adjusting the weight parameter to achieve an optimal compromise solution. It is evident that finding the right balance between reducing voltage variation and limiting the increase in LVR tap operations is crucial. In the case of this system, a weight value of $w_\mathrm{\theta}$ (= 0.05) provides a favorable trade-off, effectively minimizing voltage variation while limiting the increase in LVR tap operations.

\subsection{Supervisory Dispatch}
Figure~\ref{fig13} shows the optimal Q-V trajectories without the application of time segmentation. Notably, these trajectories deviate from the standard VV-C, with the optimal Volt-VAR points predominantly distributed in a vertical manner rather than horizontally. Thus, many of these points reside within the dead band region of the standard VV-C. To achieve a more accurate fit, we use the proposed time segmentation and focus on Segment~2 as depicted in Fig.~\ref{fig6}. This particular segment deserves attention as it corresponds to the time period characterized by large voltage variations.

We proceeded to examine the impact of varying VV-C update frequencies for the DVC. We conducted tests using update rates of 30, 60, 120, and 240 minutes. Figure~\ref{fig14} presents a sample of the optimal Q-V trajectories and the VV-Cs fitted using the two proposed approaches: \textit{curve shifting} and \textit{fitted VV-C}. These particular results focus on the use of 120-minute updates. The findings reveal an enhancement in curve fitting as there is a closer alignment between the adjusted VV-Cs and the optimal Q-V curves. Note that the fitted VV-C exhibits a closer proximity to the optimal Q-V trajectories, primarily because we have the capability to adjust the slope in this case.

We conducted additional simulations on the sample system with the DVC using the revised VV-Cs under local control. Tables~\ref{tab8} and \ref{tab9} show the statistics for voltage variation and voltage regulator operation for different cases: base case, standard VV-C, shifted VV-C, and fitted VV-C, respectively. These results demonstrate a substantial reduction in voltage variations compared to the standard VV-C when using the revised curves. Comparing these new statistics with those obtained from optimal dispatch in Table~\ref{tab6}, we observe that the improvement in reducing voltage variation is not as significant as with optimal dispatch. However, it is still notably more effective than applying the standard VV-C.

\begin{table}[b]
\vspace{-0.2in}
	\begin{center}
		\caption{\textbf{Voltage Violation in Sunny and Cloudy Days.}}    \vspace{-0.1in}
		\label{tab10}
		\centerline{\includegraphics[width=\linewidth]{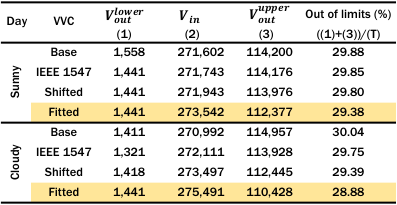}}
	\end{center}
\end{table}

\vspace{-0.1in}
\subsection{Sunny vs. Cloudy Days}
The impact of PV output variability on voltage variation is more pronounced on cloudy days compared to sunny days. Figure~\ref{fig15} presents the normalized load and PV profiles for both sunny and cloudy days. We examined the effectiveness of the DVC in mitigating high voltage variations caused by cloud cover. For this analysis, we employed a 120-minute update frequency, which demonstrated the best performance according to Tables~\ref{tab8} and \ref{tab9}. The total voltage points (T) monitored on both sunny and cloudy days are 387,360, respectively. The main observations from the simulation analysis can be summarized as follows:
\begin{table}[t]
    \vspace{-0.1in}
	\begin{center}
		\caption{\textbf{LTC and LVR Tap Changes in Sunny and Cloudy Days.}}  \vspace{-0.1in}
		\label{tab11}
		\centerline{\includegraphics[width=\linewidth]{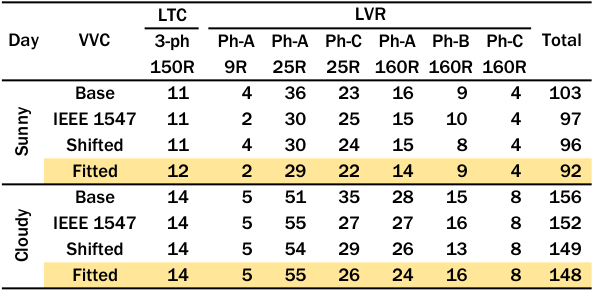}}
    \vspace{-0.2in}
	\end{center}
\end{table}

\begin{itemize}
    \item The DVC shows greater effectiveness in reducing voltage variations on cloudy days compared to sunny days due to its rapid response to PV variability. Table~\ref{tab10} demonstrates the performance of the DVC with fitted VV-C, showing a 1.7\% reduction in voltage variations on sunny day and a 3.9\% reduction on cloudy day when compared to the base case without DVC.   
    \item The proposed local dispatch schemes, namely the shifted and fitted VV-Cs, outperform the standard VV-C (i.e., IEEE Std. 1547). On sunny day, the shifted VV-C reduced voltage variations by 0.2\%, while the fitted VV-C mitigated them by 1.6\%. Similarly, on cloudy day, the shifted VV-C reduced voltage variations by 1.2\%, while the fitted VV-C achieved a greater reduction of 2.9\%.  
    \item The proposed scheme also effectively limits the increase in LVR operations. According to Table~\ref{tab11}, the DVC with the fitted VV-C reduces voltage regulator operations from 97 to 92 (5.2\% reduction) on sunny days and from 152 to 148 (2.6\% reduction) on cloudy days, respectively.
\end{itemize}

\section{Conclusion} 
This paper proposes a practical dispatching scheme designed to mitigate the rapid voltage variations caused by PV intermittency on a feeder. The proposed supervisory dispatch scheme adjusts the VV-C utilized by the local DVC controller, overcoming the limitations of existing methods. Through simulations conducted on a sample distribution feeder, the effectiveness of the proposed scheme is demonstrated. The simulations clearly indicate that using standard Volt-VAR curves for local DVC control may not effectively reduce voltage variations. 

The paper highlights the significance of the proposed approach, which employs a supervisory dispatching scheme to modify these curves, ensuring that the DVC provides efficient voltage variation reduction while minimizing LVR tap operations. Additionally, the paper emphasizes the necessity of an optimal dispatching scheme to properly modify the VV-C. The case study demonstrates the need for adjusting the VV-C about every two hours, particularly during periods of high and variable PV output. Furthermore, the optimal dispatching scheme can be used to determine the optimal DVC placement on a distribution feeder with high PV generation. The case study results illustrate that the proposed heuristics-based scheme is highly effective in determining suitable candidate locations, while maintaining computational efficiency.

\ifCLASSOPTIONcaptionsoff
\newpage
\fi 
\bibliographystyle{IEEEtran}
\bibliography{IEEEabrv,MyRefs}

\EOD

\end{document}